\documentclass[manuscript]{aastex}

\usepackage{amsmath}
\usepackage{amssymb}
\usepackage{graphicx}
\usepackage{mathrsfs}
\usepackage{bm}
\usepackage{wasysym}
\usepackage{placeins}
\usepackage{lineno}
\usepackage{array}

\shorttitle{Food Selection}

\begin{document}


\linenumbers 
\modulolinenumbers[2]

\title{Functional morphology, stable isotopes, and human evolution:\\ a model of consilience}

\author{\textbf{Justin D. Yeakel${}^1$,~Nathaniel J. Dominy${}^2$,~Paul L. Koch${}^3$,~\&~Marc Mangel${}^4$}}

\affil{${}^1$Department of Ecology and Evolutionary Biology,\\ 
University of California Santa Cruz,\\
1156 High St., Santa Cruz, CA 95064, USA\\
jdyeakel@gmail.com}

\affil{${}^2$Department of Anthropology,\\ 
Dartmouth College, Hanover, NH 03755, USA}

\affil{${}^3$Department of Earth and Planetary Sciences \\
University of California Santa Cruz, \\
1156 High St., Santa Cruz, CA 95064, USA}

\affil{${}^4$Center for Stock Assessment and Research \& \\
Department of Applied Mathematics and Statistics,\\ 
University of California Santa Cruz, Santa Cruz, CA 95064, \\
USA and
Department of Biology, University of Bergen, Bergen 5020, Norway}

\newpage

{\sf \textbf{
Foraging is constrained by the energy within resources and the mechanics of acquisition and assimilation.
Thick molar enamel, a character trait differentiating hominins from African apes, is predicted to mitigate the mechanical costs of chewing obdurate foods.
The classic expression of hyperthick enamel together with relatively massive molars, termed megadontia, is most evident in {\it Paranthropus}, a lineage of hominins that lived ca. 2.7 to 1.2 million years ago.
Among contemporary primates, thicker molar enamel corresponds with the consumption of stiffer, deformation-resistant foods, possibly because thicker enamel can better resist cracking under high compressive loads.
Accordingly, plant underground storage organs (USOs) are thought to be a central food resource for hominins such as {\it Paranthropus} due to their abundance, isotopic composition, and mechanical properties.
Here, we present a process-based model to investigate foraging constraints as a function of energetic demands and enamel wear among human ancestors.
Our framework allows us to determine the fitness benefits of megadontia, and to explore under what conditions stiff foods such as USOs are predicted to be chosen as fallback, rather than preferred, resources.
Our model predictions bring consilience to the noted disparity between functional interpretations of megadontia and microwear evidence, particularly with respect to {\it Paranthropus boisei}.
}}


\newpage

\section*{\Large  \sf{\emph{Introduction}}}

All animals must acquire and deliver food to their digestive systems. 
The mechanics of this process can result in the gradual wear, or senescence, of anatomical structures such as claws, beaks, and teeth. 
Such wear is detrimental to the foraging efficiency and reproductive success of a wide range of animals \citep{Swennen:1983hfa,Raupp:1985ub,Juanes:1992uv,Juanes:1995jx,King:2005kj,Roitberg:2005iv}. 
For mammals, the oral processing, or chewing, of food is a necessary wear-inducing behavior \citep{Stirling:1969vf,McArthur:1988kd,Skogland:1988wt,PerezBarberia:1998gj}, and natural selection is predicted to favor dental attributes that prolong chewing efficiency.
Accordingly, considerable attention has been focused on the microstructure, morphology, and functional ecology of mammalian molars, particularly the enamel.

Enamel is a hard, mineralized tissue covering the elastic and vascularized dentin, and rooted by cementum to the jaws of most mammals \citep{Lucas:2004wz}.
Oral comminution of food prior to digestion is, today, a uniquely mammalian behavior \citep{Lucas:2004wz}, although gizzards and pharyngeal teeth serve this function in birds and some teleosts, respectively, and some ornithischian dinosaurs did comminute food orally \citep{Weishampel:2004tpa}.
Some mammals have ever-growing teeth, but primates replace their molar teeth only once, after which they face an adult life of wear and occasional catastrophic damage \citep{Constantino:2010et}. 
Thus, adult primates must find a balance between the opposing advantages of enamel preservation and the consumption of foods with different propensities for enamel wear. 
In this vein, the identification of contemporary form-function relationships between tooth enamel and diet have been instructive for inferring foraging behavior in the fossil record, and dental enamel has long informed debate in the discipline of paleoanthropology \citep{Ungar:2011em}.

For example, among living mammals, relatively thick molar enamel is widely associated with the consumption of stiff, deformation-resistant (hard) foods, and it follows that hominins such as the genus {\it Paranthropus}, which possessed large `megadont' molars with hyperthick enamel \citep{McHenry:1988tb}, also consumed such foods \citep{Kay:1981tr,Kay:1985p987,Wood:2007cd,Lucas:2008p1237,Lucas:2008bd,Vogel:2008ha,McGraw:2012dj,McGraw:2012fl,Wood:2012ct,Constantino:2013ut}.
However, the identity of these stiff food objects has long puzzled researchers and fueled hypotheses on the cost of dietary specialization \citep{Balter:2012wt}.
More recently, isotopic data from a range of hominin taxa, including {\it Paranthropus}, that reveal a heavy dependence on $\rm C_4$-photosynthetic plants (which have tissues enriched in $\rm {}^{13}C$ and include tropical grasses and sedges) or possibly animals that consumed these plants \citep{Sponheimer:1999p1389,Ungar:2011em}.
Depending on the tissue, $\rm C_4$ plants can be highly resistant to fracture or deformation, with mechanical attributes that are expected to induce wear or chip the enamel of molar teeth.
Indeed, the molars of {\it Paranthropus boisei} are often heavily worn and deeply chipped \citep{Constantino:2010et}; and yet, paradoxically, the occlusal surfaces of nine well-preserved molars evince a microwear fabric that indicates a diet of soft, pliable foods \citep{Ungar:2008tj,Ungar:2010hk,Ungar:2012bp}.
These discrepant lines of evidence - indicating a diet of $\rm C_4$ foods that are simultaneously hard and soft - have been challenging to reconcile, and it is sometimes referred to as the ``$\rm C_4$ conundrum".

For {\it Paranthropus}, the consumption of $\rm {}^{13}C$-enriched tissues might have included graminivorous insects and/or the leaves, seeds, and underground storage organs (USOs) of grasses and sedges \citep{Sponheimer:2003p1378,Sponheimer:2005p1377,Yeakel:2007p1410,Cerling:2011wh,LeeThorp:2011ia}.
Recently, the USOs - bulbs, corms, and rhizomes - of grasses and sedges have attracted special attention \citep{Dominy:2012bi,LeeThorp:2012ea} because they are widespread in many savanna habitats and a central food resource for some populations of baboons and humans \citep{Post:1982bt,Barton:1993wc,Youngblood:2004it,Alberts:2005ug}.
Corms in particular are stiff and deformation-resistant \citep{Dominy:2008p864} and a significant cause of tooth wear among savanna baboons \citep{Galbany:2011cm}.
However, the mechanical and nutritional properties of these potential foods, as well as the anatomical constraints of hominin dentition, are seldom factored into interpretations of hominin foraging behavior, and the diet of {\it Paranthropus} remains obscured by disquieting discrepancies \citep{Grine:2012gw}. 
Here we attempt to bring consilience to these discrepancies using a modeling framework.

The physiological and behavioral processes that yield inconsistent interpretations of diet can be explored with foraging models that depend on the anatomical and energetic states of potential foragers.
Here we present a Stochastic Dynamic Programming (SDP) model \citep{Mangel:1988uaa,Mangel:1992wo,Houston:1999va,Clark:2000tra} to quantify the optimal foraging decisions for organisms that must balance energetic gain with enamel wear, while accounting for the stochastic effects of a variable environment. 
We base our model on measurements for anthropoid primates and focus specifically on decisions affecting hominins in savanna-woodland environments. 
We determine decision matrices in which specific food resources are chosen to maximize an animal's fitness conditional on two internal states: its energetic reserves and molar enamel volume.

This model-based approach is well-suited to test a variety of important questions about the effects of dental enamel on foraging, and we focus on three potentially informative lines of inquiry. 
First, and most essential, how does the quantity of enamel  influence the foraging strategies of savanna-woodland anthropoid primates, and how does megadont dentition alter these strategies? 
Second, to what degree do these foraging decisions depend on resource quality and quantity, where the quality and quantity of particular food items can vary depending on the environmental conditions? 
Third, can extradentary mechanical advantages, such as peeling, pounding, or cooking alter the influence of dental enamel, and to what extent do these alterations provide fitness benefits? 
Finally, we relate our model predictions to paleontological evidence of hominin diets, and conclude by showing that the model framework presented here can be used to both predict and inform paleodietary data.

\section*{\Large  \sf{\emph{Models and Analysis}}}

Models based on Stochastic Dynamic Programming are recognized as one of the best ways of predicting the evolutionary end-points for natural selection.  In this section, we outline the structure of the SDP model from which we determine fitness-maximizing foraging behaviors of hominin species.
First, we define energetic state and enamel volume as the state variables of the model, and describe the processes that govern how these state variables change over time.
We also introduce three factors that influence an organism's state:
1) the probability of finding different amounts of food (including not finding it),
2) the probability of losing a given amount of enamel as a function of chewing different foods, and
3) the quality of the environment at a given time.
Second, we introduce the fitness function, which depends upon the state of the organism and time.  Starting at a fixed final time, we show one can iterate the fitness function backwards in time, thus determining both fitness at earlier times and foraging decisions (the decision matrix) as a function of state.  
As the current time moves further and further from the final time, the decision matrix becomes independent of time (stationary), only depending upon physiological state.  
Third, we combine the stationary decision matrices with a Monte Carlo simulation going forward in time (forward-iteration) to examine the consequences of different foraging behaviors as a function of an organism's anatomical attributes and/or its ability to externally modify its food.

\subsection*{\small\sf{\textbf{STATE DYNAMICS}}}

We model the foraging decisions of an organism as a function of two principle state variables: 
1) its net energy reserves at time $t$, $X(t) = x$, and
2) its enamel volume at time $t$, $V(t) = v$, where time is measured in days.
We model a single unit of energy as 10 Megajoules [MJ], equivalent to 2388 kcal and roughly equal to the energy in 1 kg of animal tissue \citep[Wolfram][]{WolframResearchInc:2010ub}.
Accordingly, the maximum potential energy reserves for an organism, $x_{\rm max}$, is its body size, such that $x_{\rm max} = 70$ for a 70 kg organism. 
A unit of molar enamel volume $v$ corresponds to 100 ${\rm mm}^3$. 
Specific properties of molar anatomy correlate with body size \citep{Shellis:1998fp}, and we use these relationships to approximate maximal (i.e. unworn) molar enamel volume, $v_{\rm max}$ as a function of $x_{\rm max}$, for both non-megadonts and megadonts (see Appendix S1, Figure S1).
Both the potential energy gained from food and its impact on an organism's enamel change as a function of food mechanical properties.
We consider an approximating measurement for the mechanical properties of food taking into account both the elasticity (Young's modulus, $E_i$, [MPa]) and the fracture toughness ($R_i$, [Jm$^{-2}$]) of food $i$, which approximates `hardness', measured as $\sqrt{E_i R_i}$ \citep{Lucas:2008bd}.
We let $\eta_i$ denote the digestibility of food $i$ ranging between $\eta_i=0$ (indigestible) and $\eta_i=1$ (completely digestible)  \citep[\emph{sensu}][]{Lucas:2000bca}.
We assume that an individual dies when its energy reserves fall below $x_{\rm crit} = (3/4)x_{\rm max}$ or its enamel volume falls below $v_{\rm crit}$ (see Appendix S1).

We let $\gamma_i$ (in units of $x$) denote an organism's energetic gains for food type $i$ and let $\eta_i$ denote the digestibility of the food (Table 1).
Because larger animals gain relatively more calories per foraging bout, energetic gain is calculated as $\gamma_i = ({\rm energy~density}/2388)\cdot(x_{\rm max}/10)$, where the constant (1/2388) normalizes the energy density of foods to units of $x$, and the modifier $(x_{\rm max}/10)$ ensures that gain scales weakly with body size.
We assume that foraging behavior is primarily governed by caloric, or energetic, limitations \citep{Rothman:2011kh}, and model the daily cost of foraging for food type $i$, $c_i$ (in units of $x$), as a function of an organism's body size, and the aggregation of food on the landscape.
We modified the estimates of daily energetic expenditure (kcal/day) by \citet{Leonard:1997usa} to model daily energetic cost, such that $c_i = C_1\cdot {\rm RMR}\cdot(1/\xi_i)$/2388, and Resting Metabolic Rate is ${\rm RMR} = 69.1x_{\rm max}^{0.761}$, where $C_1$ is the activity constant ($C_1=3.80$ for moderate activity), the constant (1/2388) operates as before, and $\xi_i$ is the mean encounter rate for food $i$, such that $(1/\xi_i)$ is proportional to foraging time.
Foods that are encountered more frequently (high $\xi_i$) thus have lower per encounter foraging costs.
We assessed a costlier version of the model, where $c_i = (C_1\cdot {\rm RMR}\cdot(1/\xi_i) + C_2\cdot{\rm RMR})$/2388, where $C_2=1.2$, accounting for additional daily costs independent of food choice \citep{Leonard:1997usa}.

We identify four general food groups:
1) a nutritious, mechanically pliable, patchily distributed food (e.g. fruit),
2) a non-nutritious, mechanically hard, widely distributed food (e.g. leaves from ${\rm C_4}$-photosynthetic grasses),
3) a nutritious, mechanically hard, widely distributed food (e.g. USOs), and
4) a highly-nutritious, potentially hard, patchily distributed food (e.g. arthropods or more generally small quantities of animal tissue).
We set the food energy density to be 717, 150, 785, and 1518 kcal/kg for fruit, grass leaves, USOs, and arthropods/animal tissue respectively \citep[Wolfram][]{WolframResearchInc:2010ub}. 
The mechanical properties of the food groups are measured by toughness [Jm$^{-2}$]: $R=(561,330, 265,1345)$, and Young's modulus [MPa]; these are  $E = (1,10,5,200)$ for fruits, grass leaves, USOs, and arthropods with fracture-resistant exoskeletons, respectively \citep{Lucas:2004wz,Williams:2005dsa,Dominy:2008p864,Yamashita:2009fa}.
 We used a conservatively low value for the fracture toughness of grass leaves in our model \cite[$330~{\rm Jm^{-2}}$;][]{Lucas:2004wz}.
Although the fracture toughness of East African grasses is typically $> 1000~{\rm Jm^{-2}}$ (N.J. Dominy, unpublished data), we assume that a grazing primate with bunodont molars would selectively consume tender grass leaves.

Many primates are known to modify the mechanical properties of foods before they are consumed \citep{Altmann:2009el}.
We consider four extradentary processing capabilities: 
1) none, where the mechanical properties are as described,
2) peeling, pounding, or cooking USOs, \citep[$R_{\rm USO}=138$ and $E_{\rm USO}=5$;][]{Dominy:2008p864},
3) peeling arthropod exoskeletons ($R_{\rm arthropod}=306$ and $E_{\rm arthropod}=0.22$) (alternatively, this mechanical advantage can account for swallowing arthropods with minimal chewing), and 
4) a combination of mechanically altering both USOs and arthopods.

The energetic and enamel state of an organism change over time, and these changes are due to both deterministic and stochastic processes.
The energetic state of an organism depends primarily on the number of resources found and the amount of energy spent in a given foraging period.
We let the random variable $K$ represent the number of food items found in a single foraging period and that with probability $f_K(k)_i$ an individual finds $k$ items of food type $i$. In Appendix S2, we derive the negative binomial model used for food encounters. 
We maintain this notation, upper case for random variables and lower case for specific values, for all stochastic variables.
Because an organism's daily consumption is limited by gut volume, daily caloric gain is bounded by $x_s = (1/5)\cdot x_{\rm max}$ \citep[proportional to average anthropoid \% gut volume;][]{Milton:1989vt}.
Thus, if $k$ items of food type $i$ are found in period $t$

\begin{equation}
 X(t+1) = X(t) + {\rm min}(k\eta_i\gamma_i,x_s) - c_i.
 \label{energy}
\end{equation}

Enamel volume decreases as an animal consumes resources.
Although the underlying mechanisms of enamel loss are poorly understood \citep{Lucas:2008p1237}, siliceous particulate matter is probably the most significant cause of abrasion \citep{Lucas:2012gj}.
We assume that hard foods (high $\sqrt{E_iR_i}$ values) promote increased use of the dentition \citep[cf.][]{Organ:2011td}, and that such use induces wear regardless of the specific cause.
We set enamel wear, $\Delta v(\Omega)$, to be a function of: 
1) the mechanical properties of food $i$ and 
2) a stochastic decrease in enamel volume (determined by $\Omega$).
Because enamel is a nonrenewable resource, this wear cannot be undone.
\citet*{Teaford:1989bc} showed that the consumption time for vervet monkeys ({\it Chlorocebus}) that ate a diet of raw Purina monkey chow was 8x greater than that for vervets fed on pre-mashed monkey chow.
With respect to enamel wear, this is equivalent to chewing 8x as much food.
Teaford and Oyen also showed that the enamel thickness decreased by ca. $0.58~\mu {\rm m~day}^{-1}$ when vervets fed on the raw diet, versus ca. $0.24~\mu {\rm m~day}^{-1}$ when they fed on the pre-mashed diet.
We assumed a linear relationship between the loss of enamel thickness \citep[from][]{Teaford:1989bc}, and consumption time, or, alternatively, the amount of food consumed, $k$ (with a slope $b = 0.0425$).
The lower-bound of this relationship ($\bar{\omega}= 0.24~\mu{\rm m}$) represents the expected basal enamel wear that occurs irrespective of consumption, and we used it to parameterize the stochastic variable $\Omega$.
Accordingly, given that $A$ is the molar enamel surface area and $E_{MC}$ and $R_{MC}$ are scaling constants denoting the average Young's modulus (50.44 MPa) and fracture toughness (1030.55 Jm${}^{-2}$) of monkey chow, respectively \citep{Williams:2005dsa}, when $k$ items of food type $i$ are found in period $t$

\begin{equation}
V(t+1) = V(t) - \underbrace{\frac{A}{250}\left(\frac{bk \sqrt{E_iR_i}}{{\sqrt{E_{MC}R_{MC}}}} + \Omega \right)}_{\Delta v(\Omega)}.
\label{wear}
\end{equation}

\noindent The constant 1/250 scales tooth wear to ensure the organism attains its expected longevity \citep{Lindstedt:1981p1038}, and accounts for
1) overestimation of molar enamel area (our allometric estimation includes the lateral aspects of molar surfaces), and
2) the notion that wear is a complex action affecting a small fraction of the occlusal surface at a given time \citep{Lucas:2004wz}.

The basal loss of enamel thickness has an expected value ${\rm E}(\Omega)=\bar{\omega}= 0.24~\mu{\rm m}$. 
As such, chewing and the daily wear unassociated with chewing have variable effects on enamel wear \citep{Lucas:2004wz} .
Specifically, enamel wear is typically small, but occasionally large, and is realized when the organism chips or looses a tooth or part of a tooth \citep[cf.][]{Boccara:2004p2506}.
To capture this property, we model the probability that $\Omega$ falls within the small interval $\omega$ and $\omega + {\rm d}\omega$,  $f_\Omega(\omega)$, by a lognormal distribution, where ${\rm E}(\Omega)=\bar{\omega}$ and ${\rm Var}(\Omega)=\sigma^2$ (see table 1).

%
%

Finally, we introduce changing habitat quality as a stochastic environmental variable that affects both the nutritional gains and foraging costs of foods at a given time.
Habitat quality can be rich ($Q(t)=\rm r$) or poor ($Q(t) = \rm p$) at time $t$, and changes through time according to a transition probability matrix ${\bm \rho} = (\rho_{\rm rr}, \rho_{\rm rp}; \rho_{\rm pr}, \rho_{\rm pp}$), where - for example - $\rho_{\rm rp}$ is the probability of transitioning from a rich quality habitat at time $t$ to a poor quality habitat at time $t+1$.
Changes in habitat quality alter energetic gain, the mean encounter rate, and the dispersion of different foods.
We set energetic gain to decrease by 10\% in poor quality habitats relative to rich quality habitats.
Moreover, the mean encounter rate ($\xi_i$) as well as the dispersion of food ($\nu_i$) are modified by $Q(t)$, such that food resources are more easily found (higher $\xi_i$) and are less patchily distributed (higher $\nu_i$) in rich quality habitats (see Appendix S2 for a detailed derivation of dispersion and encounter rates of foods).
USOs are stored underground and have evolved to retain high nutrient loads during periods of environmental stress \citep{Copeland:2004to}.
We incorporate this quality by holding the energetic gain, encounter rate, and dispersion of USOs constant, irrespective of habitat quality.


With this basic framework, we assess the influence of `wet', `dry', and `autocorrelated' environmental conditions on foraging decisions.
Wet environments have high values of $\rho_{\rm rr},~\rho_{\rm pr}$, and low values of $\rho_{\rm rp},~\rho_{\rm pp}$ (such that habitat quality is generally rich), whereas dry environments are the opposite.
Autocorrelated environments are unlikely to change from their current state, and thus have high values of $\rho_{\rm rr},~\rho_{\rm pp}$, and low values of $\rho_{\rm rp},~\rho_{\rm pr}$ (see table 1).
We recognize that natural systems are more idiosyncratic, however this simplification allows us to assess the effects of changing habitat quality over time with minimal added complexity.

\subsection*{\small\sf{\textbf{MAXIMIZING FITNESS BY FOOD CHOICE}}}

We consider a nonbreeding interval of length $T$ during which only foraging decisions influence fitness.
This interval will ultimately become sufficiently large (see below) so that we can use decisions that are independent of time \citep[cf.][]{Mangel:1988uaa,Clark:2000tra}, and assume that at the end of this interval, the fitness of an individual with energy reserves $x$ and enamel volume $v$ is $\Phi(x,v)$.  
For numerical computations we use


\begin{align}
\Phi(x,v) &= \frac{1}{2}\left(2- \frac{x_{\rm crit}}{x} -  \frac{v_{\rm crit}}{v}\right), ~~~~~\mbox{where}~ \left\{
	\begin{array}{l l}
	x > x_{\rm crit}\\
	v > v_{\rm crit}
	\end{array}, \right. \nonumber \\
\Phi(x,v) &= 0, ~~~~~~~~~~~~~~~~~~~~~~~~~~~~~~\mbox{otherwise}.
\label{Eq:TermFit}
\end{align}

\noindent The maximum fitness at $t=T$ is realized by an organism with $X(T)=x_{\rm max}$ and $V(T)=v_{\rm max}$, and the rate of fitness decline increases as $x$ and $v$ approach $x_{\rm crit}$ and $v_{\rm crit}$.
We explored alternatives such as $\Phi(x,v)=(1-x_{\rm crit}x^{-1})(1-v_{\rm crit}v^{-1})$ and they had little effect on the qualitative predictions.  
We scaled the terminal fitness function to be 1, so it is easiest to consider it as survival after $T$ for an individual whose end state is $X(T)=x, V(T)=v$.

We assume that natural selection has acted on behavioral decisions concerning diet (food choice) conditioned on energetic state, enamel volume, and the probability of transitioning from rich or poor habitat quality. We define fitness functions

\begin{subequations}
\begin{align}
 F_{\rm r}(x,v,t) &=  \underset{i}{\rm max}~{\rm E}\left\{ \Phi(X(T),V(T))|X(t)=x,V(t)=v, Q(t)=\mbox{r}\right\} \label{eq_fit1}, \\
 F_{\rm p}(x,v,t) &=  \underset{i}{\rm max}~{\rm E}\left\{ \Phi(X(T),V(T))|X(t)=x,V(t)=v, Q(t)=\mbox{p}\right\},\label{eq_fit2}
\end{align}
\end{subequations}

\noindent where the maximization over $i$ chooses the food that maximizes fitness given energy reserves, enamel volume, and habitat quality.  
By definition, at time $T$

\begin{equation}
F_{\rm r}(x,v,T) = F_{\rm p}(x,v,T) = \Phi(x,v). \nonumber
\end{equation}


For time periods before the terminal time $t=T$, an organism must survive mortality independent of starvation or enamel loss and choose the fitness maximizing food, given the stochasticity in food encounter.  
If the probability of death in a single period is set to $m$ \citep [$ \approx {\rm e}^{-10}$ or $4.5\times10^{-5}$, estimated for a subadult male chimpanzee; cf.][]{Bronikowski:2011jd}, then $F_{\rm r}(x,v,T)$ and  $F_{\rm p}(x,v,T)$ satisfy the equations of Stochastic Dynamic Programming  (SDP), such that

\begin{subequations}
\begin{align}
F_{\rm r}(x,v,t) =
&\underset{i}{\rm max}~(1-m)\Bigg( \sum_{k=0}^{k_{\rm max}}f(k_{\rm r})_i \Big( \rho_{\rm rr}{\rm E}_\Omega \{F_{\rm r}(x_{\rm r} + {\rm min}(k\eta_i\gamma_i,x_s)_{\rm r} - (c_i)_{\rm r}, v - \Delta v(\Omega), t+1)\} \nonumber \\  
& + \rho_{\rm rp}{\rm E}_\Omega \{F_{\rm p}(x_{\rm r} + {\rm min}(k\eta_i\gamma_i,x_s)_{\rm r} - (c_i)_{\rm r}, v - \Delta v(\Omega), t+1)\} \Big)\Bigg), \label{Eq_sdp1} \\
F_{\rm p}(x,v,t) =
&\underset{i}{\rm max}~(1-m)\Bigg( \sum_{k=0}^{k_{\rm max}}f(k_{\rm p})_i \Big( \rho_{\rm pr}{\rm E}_\Omega \{F_{\rm r}(x_{\rm p} + {\rm min}(k\eta_i\gamma_i,x_s)_{\rm p} - (c_i)_{\rm p}, v - \Delta v(\Omega), t+1)\} \nonumber \\  
& + \rho_{\rm pp}{\rm E}_\Omega \{F_{\rm p}(x_{\rm p} + {\rm min}(k\eta_i\gamma_i,x_s)_{\rm p} - (c_i)_{\rm p}, v - \Delta v(\Omega), t+1)\} \Big)\Bigg),
\label{Eq_sdp2}
\end{align}
\end{subequations}

\noindent where the expectation ${\rm E_\Omega}$ is taken with respect to the random variable $\Omega$ (equation \ref{wear}). 
These equations identify the food $i$ that maximizes fitness for given energetic reserves $X(t) = x$, enamel volume $V(t) = v$, and habitat quality $Q(t)$ at time $t$.
As equations (5) are solved backward in time, in addition to obtaining the values of fitness, we create decision matrices $D_{\rm r}(x,v,t)$ and $D_{\rm p}(x,v,t)$ characterizing the optimal choice of food in a rich or poor environment given that $X(t)=x$ and $V(t)=v$.
Thus, the two decision matrices (for rich and poor quality) depend upon the habitat quality transition matrix $\bm \rho$, but we suppress that notation for ease of reading.

As $t$ moves backwards further and further away from $T$ the fitness maximizing decisions become independent of time and depend only upon state, which accords with the intuition that far from the time at which fitness is assessed, the behavior of an organism is predicted to depend on its state and on the environment, but not on the current time.
Decisions that maximize fitness at $t\ll T$ are thus stationary with respect to time. 
We used these stationary decisions, which we denote by $D_{\rm r}^*(x,v)$ and $D_{\rm p}^*(x,v)$ for further analysis.
We confirmed stationarity by calculating the summed square differences between decision matrix solutions from $t+1$ to $t$, such that $\Delta D(t) = \sum_{v,x} (D(x,v,t+1) - D(x,v,t))^2$, for $t = T-1$ to $t \ll T$ and we assumed stationary decisions had been reached when $\Delta D(t) \rightarrow 0$ for $t \ll T$ (for an example, see Figure S2).

\subsection*{\small\sf{\textbf{FORWARD ITERATION}}}
We used a Monte Carlo simulation moving forward in time (forward iteration algorithm \citep{Mangel:1988uaa,Clark:2000tra} to assess the impact that fitness maximizing foraging decisions (given by $D^*_{\rm r}(x,v)$ and $D^*_{\rm p}(x,v)$) have on the expected future fitness of individuals by iteratively solving for the state dynamics of simulated foragers over time, given the state dynamics in equations (\ref{energy}) and (\ref{wear}).
We let  $\tau$  denote forward-iterated time units experienced by simulated individuals making foraging decisions in accordance to the stationary decision matrices, as opposed to the time units $t$ used to calculate stationary decision matrices.
At each time $\tau$,  the $n^{th}$ simulated individual, with states $X_n(\tau)$ and $V_n(\tau)$ forages for the food $i$ determined by the decision matrices $D^*(X_n(\tau),V_n(\tau)|Q(\tau))$. 

To test whether and to what extent mechanical advantages conveyed fitness benefits to hominin primates, we quantified expected future fitness, ${\hat F}(\tau|D^*,Q(\tau))$, for $n=1, 2, ..., N=100$, 50 kg individuals, with maximal foraging costs for days $\tau=1$ to $\tau_{\rm max}=10950$ (expected lifespan of 30 years) given both the stationary decision solutions and habitat quality.
As energy reserves and/or enamel volume decrease over the lifetime of an individual, ${\hat F}$ is expected to decrease similarly.
We quantified the expected future fitness at time $\tau$ of a population, 
\begin{equation}
{\hat{F}}(\tau | D^*, Q(\tau)) = \frac{1}{N} \sum_{n=1}^N F^* \big(X_n(\tau),V_n(\tau) | D^*, Q(\tau)\big).
\end{equation}
where $F^* \big(X_n(\tau),V_n(\tau) | D^*, Q(\tau)\big)$ is the optimal fitness for individual $n$ at time $\tau$ given its physiological states and the environment.  


We explored the potential adaptive benefits of megadontia and extradentary mechanical advantages using two approaches.
First, we compared the proportions of foods identified to maximize fitness in accordance to the stationary decisions $D^*_{\rm r}(x,v)$ and $D^*_{\rm p}(x,v)$.
Organisms that are predicted to utilize a particular resource across a greater proportion of states $(x,v)$ may have fitness benefits in environments where those resources are plentiful.
However, although the percentage of foraging choices in decision matrices is an efficient summary of potential dietary behavior, it should not be viewed as the proportional contribution of food to an individual's diet over time, which is calculated with the forward iteration algorithm.
To determine whether megadontia provided fitness benefits over the lifetime of an individual organism, we compared expected future fitness, ${\hat{F}}$, for populations of individuals with and without megadont dental anatomy (incorporated into the model by altering $v_{\rm max}$; see Appendix S1), mechanical advantages, and during both wet environments (where rich quality habitats are more likely) and dry environments (where poor quality habitats are more likely). 

\section*{\Large  \sf{\emph{Results}}}

\begin{figure*}[h]
   \centering
   \includegraphics[width=0.75\textwidth]{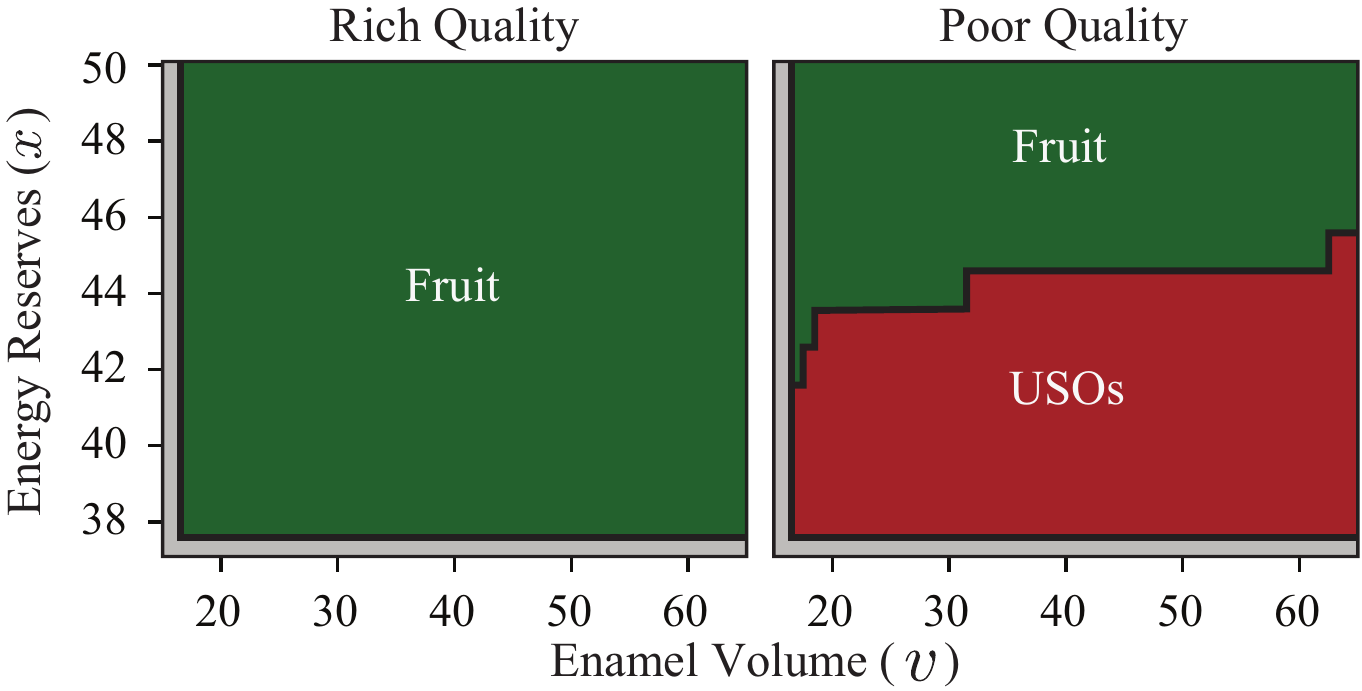}
      \caption{
      Stationary solutions to the fitness-maximizing equations $F_{\rm r}(x,v)$ (rich quality habitat) and $F_{\rm p}(x,v)$ (poor quality habitat) for a 50 kg anthropoid primate with no mechanical advantages. 
      There are no qualitative differences between wet, dry, or autocorrelated conditions. Gray elements to the left and bottom of the plots denote values of $(x,v)$ resulting in mortality.
      }
      \label{fig:DM_mn}
\end{figure*}

Based on the stationary solutions, we predict that energy reserves and enamel volume have large consequences for diet choice.
In rich quality habitats, foods with the energetic and mechanical properties of fruit maximize the fitness of animals without an extradentary mechanical advantage across all potential states $(x,v)$ (Figure \ref{fig:DM_mn}).
In poor quality habitats, such food maximizes fitness only if energy reserves are high; as reserves decline, the optimal resource shifts from fruit to plant USOs.
Plant USOs confer similar energetic gain as fruit, however we hold the mean encounter rate and dispersion of USOs constant in both rich and poor quality habitats, whereas fruits are patchier in poor habitats (see Table 1).
As enamel volume declines with age, the mechanical hardness of USOs, which produce greater enamel wear, is predicted to promote an increased reliance on riskier but mechanically pliable foods such as fruit.

Stationary decision matrices reveal that hominins with megadontia can maximize fitness by incorporating a relatively greater proportion of obdurate foods in poor quality habitats.
With no mechanical advantage, megadont decision matrices show a reduction in the percentage of fruit, and an increase in USOs relative to non-megadonts (Figure \ref{fig:Tern}).
As mechanical advantages are introduced, megadont decision matrices show similar percentages of each food item as those of non-megadonts with one important difference: regardless of the mechanical advantage, megadont decision matrices include a greater percentage of USOs.

\begin{figure*}[h]
   \centering
   \includegraphics[width=1\textwidth]{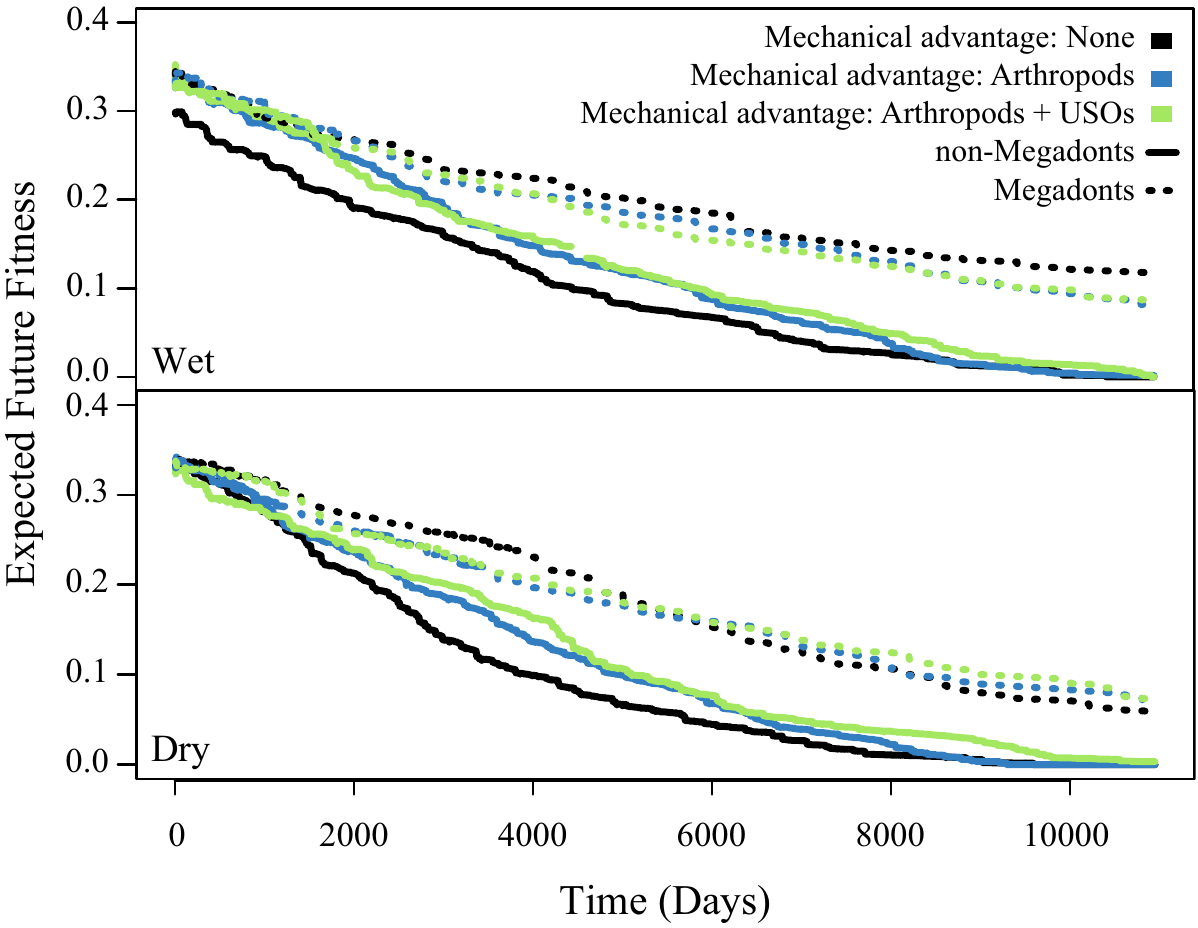}
      \caption{
      Expected future fitness trajectories for $N=100$, 50 kg non-megadont (solid) and megadont (stippled) hominins over an estimated lifespan with varying mechanical advantages (none, arthropods, arthropods + USOs), during both wet and dry environmental conditions.
      }
      \label{fig:fit}
\end{figure*}

For all simulated populations, forward iterations reveal that expected future fitness decreases sharply early in life, but saturates as the population reaches its expected lifespan of 30 years (10950 days) (Figure \ref{fig:fit}).  
This is due to wear on enamel and potential decline in energy reserves going forward in time, resulting in lower future fitness.
Because the decision matrices for the USO mechanical advantage are nearly identical to the no mechanical advantage scenario, we show only forward equation results for the latter.
Our results point to an important difference between the three mechanical advantage scenarios that are considered (none, arthropods, arthropods + USOs; Figure \ref{fig:fit}A,B; solid lines).
Both arthropod and arthropod + USO mechanical advantages appear to have large impacts on expected fitness. 
For both wet and dry environmental conditions, having either mechanical advantage provides large fitness benefits, but the difference in fitness {\it between} mechanical advantages is small, particularly when habitat quality is generally rich (wet conditions).

The fitness advantages of megadontia are more obvious.
Populations with this character trait have greater expected future fitness than those without megadontia - irrespective of mechanical advantage - and these differences are more exaggerated later in life (Figure \ref{fig:fit}A,B; stippled lines).
Moreover, the predicted fitness benefits generated by a mechanical advantage are generally less for populations with megadontia.

\begin{figure*}[h]
   \centering
   \includegraphics[width=1\textwidth]{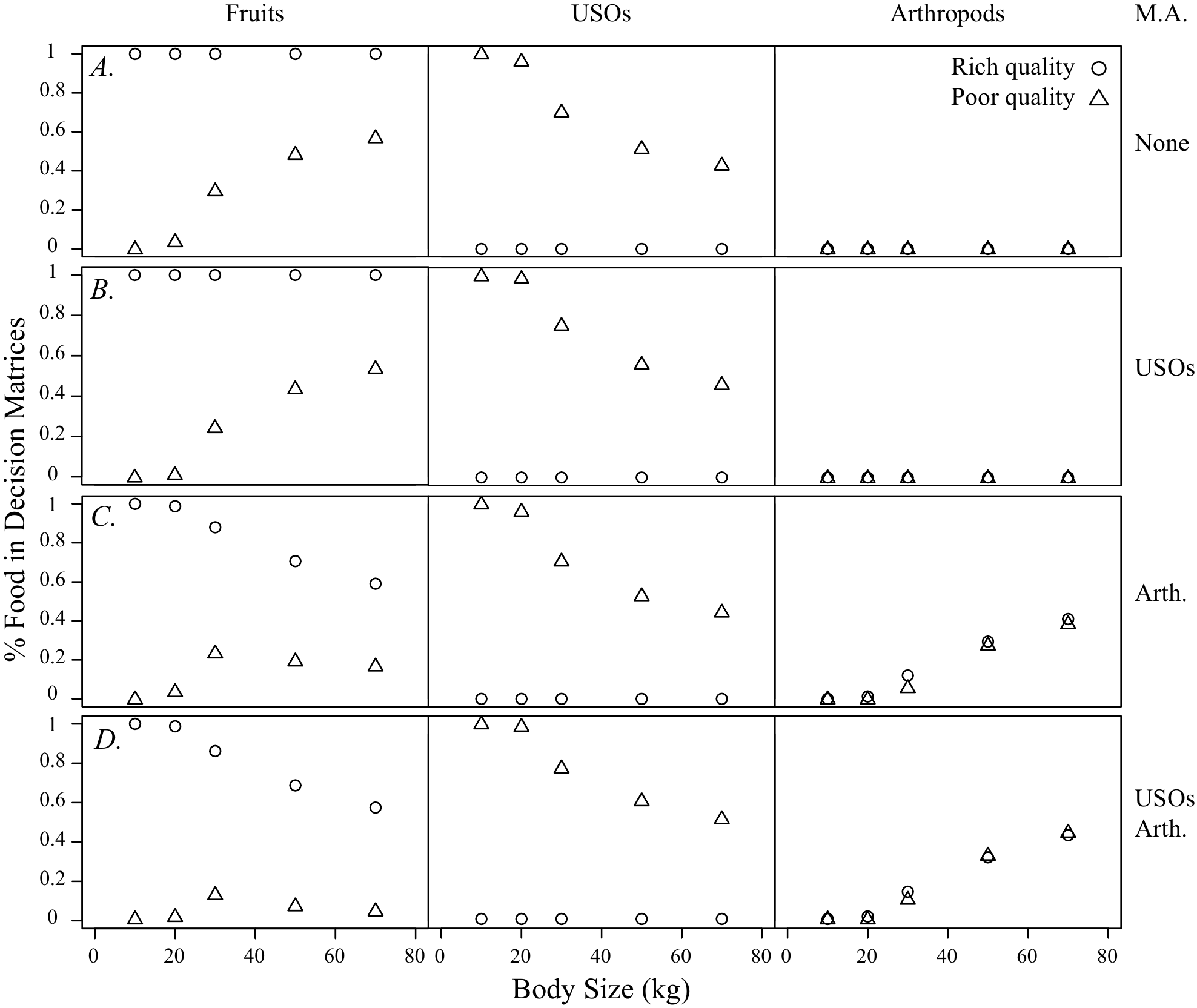}
      \caption{
      The proportional contribution of foods to the decision matrices of anthropoid primates with body sizes ranging from 10 to 70 kg. Contributions of foods for the no mechanical advantage scenario ({\it A.}), the USO advantage scenario ({\it B.}), the arthropod advantage scenario ({\it C.}), and the arthropod + USO advantage scenario ({\it D.}). 
      Grass leaves are not found to be optimal foraging solutions in any decision matrix. Results are shown for autocorrelated environmental conditions; results for wet and dry conditions were qualitatively similar.
      }
      \label{fig:PercFood}
\end{figure*}

Because foraging costs scale nonlinearly with body size, optimal foraging decisions vary accordingly. 
For larger animals and for each environmental scenario in our model (wet, dry, and autocorrelated), a poor habitat quality is strongly associated with the consumption of riskier foods with higher energetic yields such as fruit, whereas more ubiquitous foods such as USOs are an important supplement (Figure \ref{fig:PercFood}A).
Animals with smaller body sizes tend to rely on USOs exclusively.
When habitat quality is rich, both smaller- and larger-bodied animals switch to a diet of energy dense foods (fruit).
In the absence of an extradentary mechanical advantage, extremely energy dense, but relatively rare foods such as arthropods are avoided by animals of any size, regardless of habitat quality.
As body size increases, the role of plant USOs remains constant, however arthropods (highest nutritional gain and lowest probability of encounter) become favored over fruit (Figure \ref{fig:PercFood}). 
Thus, in both rich and poor quality habitats, large-bodied animals increase the percentage of risky foods if their mechanical properties can be altered to preserve enamel (Figure S3). 
Smaller-bodied animals lack the energetic reserves required to forage on rare, but energy dense foods such as arthropods, regardless of their mechanical advantages.

\begin{figure*}[h]
   \centering
   \includegraphics[width=1\textwidth]{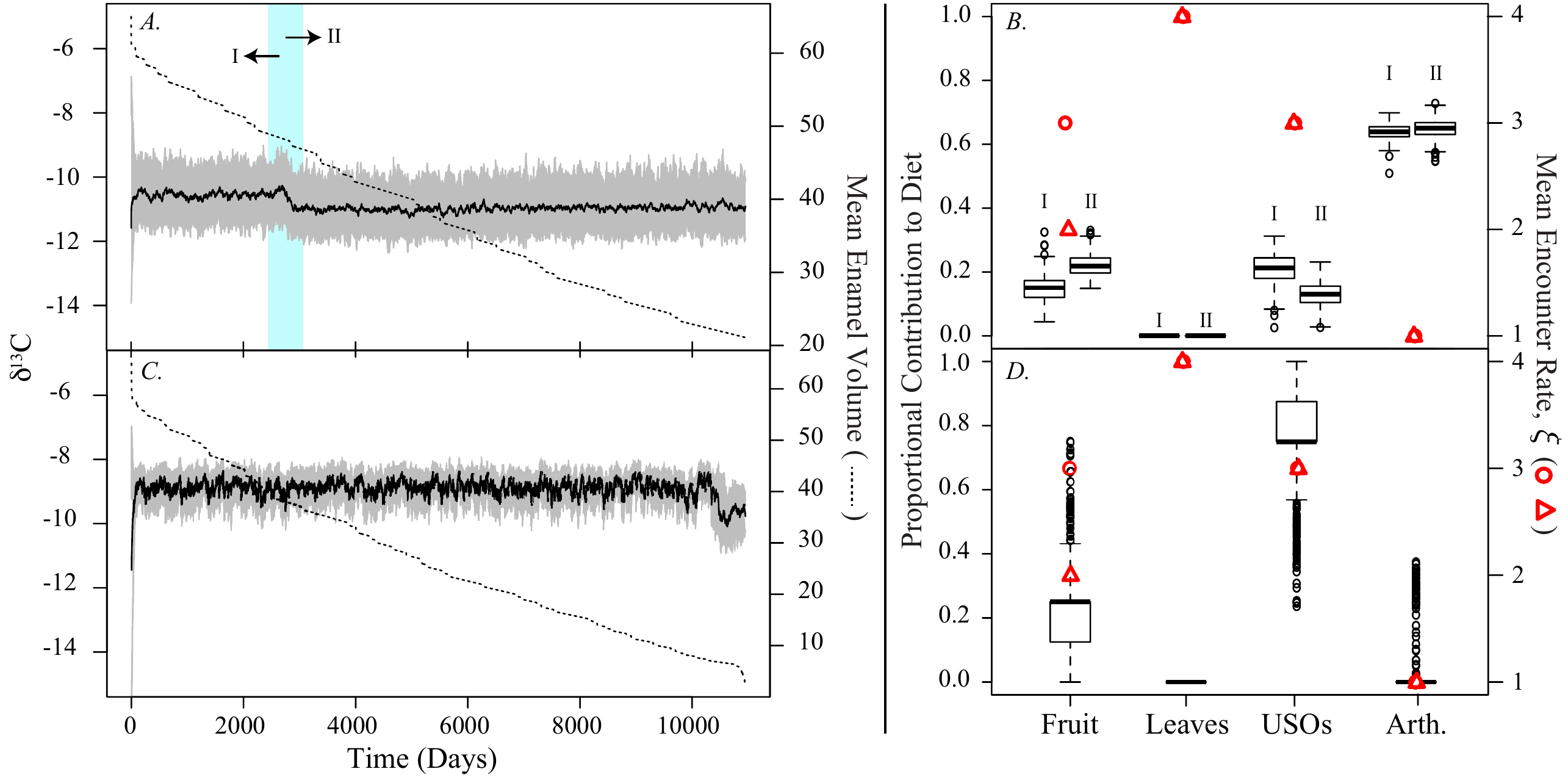}
      \caption{
            Forward simulation of the $\delta^{13}{\rm C}$ values (black line denotes running mean; gray band denotes maximum and minimum values), mean enamel volume, and the proportional contribution of food-items to the diets of $N=100$, 50 kg individuals foraging in a dry environment over an estimated lifespan. {\it A.} and {\it B.} When foraging costs are minimal, a dietary switch is observed to occur near day 3500, and labels I and II denote the pre- and post-diet switch.
            {\it C.} and {\it D.} The same simulation when foraging costs are elevated.
            In panels B. and D., the red circles and triangles denote the mean encounter rate for each food in rich quality and poor quality habitats, respectively. 
      }
      \label{fig:IsoSim}
\end{figure*}

Given that the food choices in our SDP model are associated with a distribution of $\delta^{13}{\rm C}$ values, we can use a forward iteration framework to explore how the accumulated $\delta^{13}{\rm C}$ values of  individuals within a population change over time as a function of energetic reserves, enamel volume, and the prevailing environmental conditions (see Appendix S3 for details).
Our results show that the $\delta^{13}{\rm C}$ values of a simulated population of $N=100$, 50 kg anthropoid foragers capable of mechanically altering both arthropods and USOs is influenced by both energetic reserves and enamel volume.
In dry environments and where foraging costs are minimal, the mean $\delta^{13}{\rm C}$ value of simulated foragers remains relatively high ($\delta^{13}{\rm C}_{\rm avg} \approx -10.5 \permil$; Figure \ref{fig:IsoSim}A), due to a greater reliance on USOs (Figure S3).
After day 3500, $\delta^{13}{\rm C}_{\rm avg}$ declines to $ -11.2 \permil$ as the proportional contribution of USOs decreases and that of fruits increases (Figure \ref{fig:IsoSim}B).
This highlights the increasing importance of foods that are less obdurate as enamel is worn - despite greater energetic costs - as well as the accompanying decrease in the mean $\delta^{13}{\rm C}$ value of a consumer population over its lifespan.

If foraging costs are too great, low risk, obdurate foods are preferred despite greater enamel wear, resulting in a higher $\delta^{13}{\rm C}_{\rm avg}\approx -8.8 \permil$ (Figure \ref{fig:IsoSim}C).
In this case, our model predicts $\delta^{13}{\rm C}$ values equivalent with those observed for {\it A. africanus} and {\it P. robustus} \citep{Ungar:2011em}.
In costlier environments (where energetic cost includes both foraging costs as well as daily costs independent of food choice), USOs tend to maximize fitness until late in life (Figure \ref{fig:IsoSim}D), when the cost of reduced enamel volume supersedes the risks of foraging on pliable but rare foods.

\begin{figure*}[h]
   \centering
   \includegraphics[width=1\textwidth]{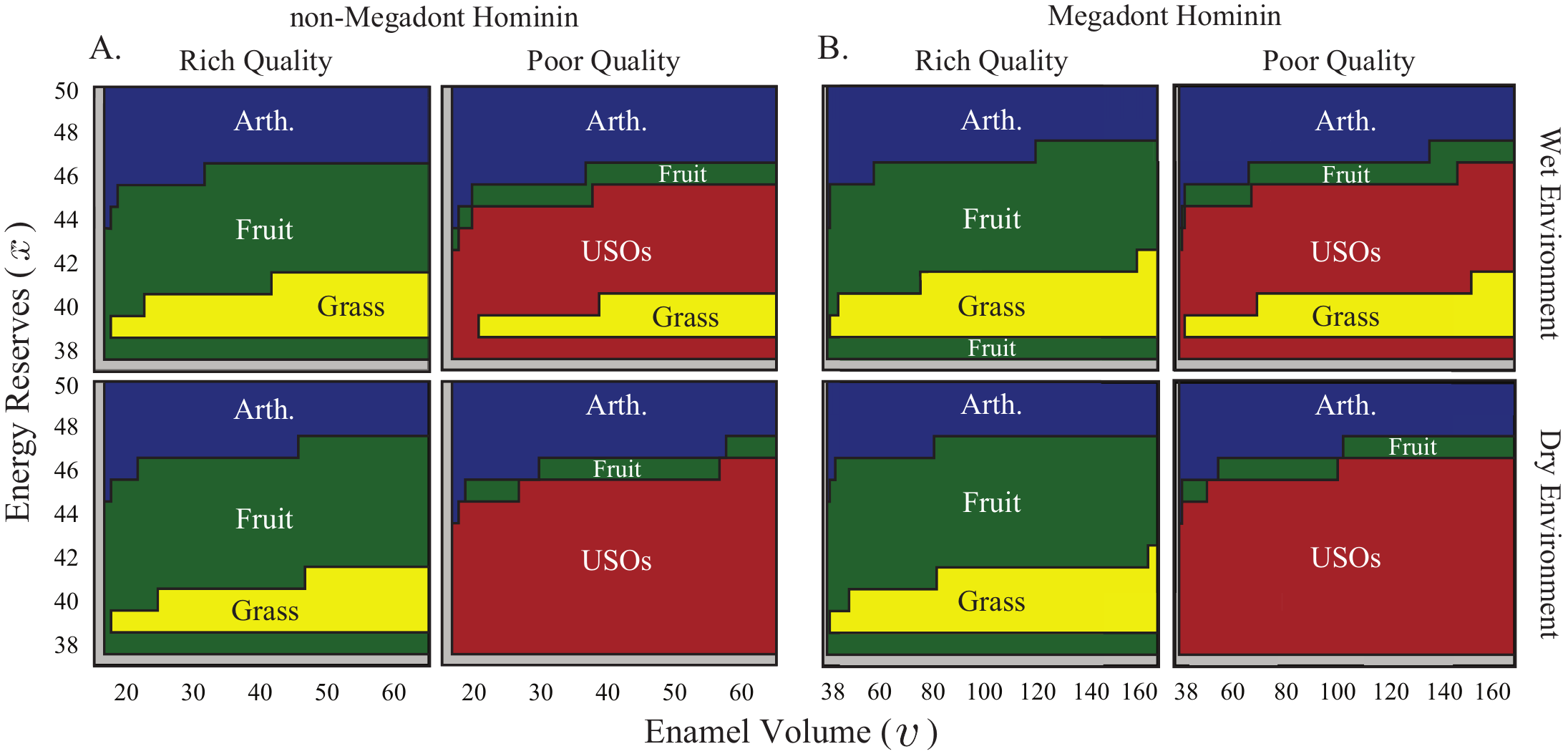}
      \caption{
      Stationary solutions for the fitness-maximizing equations, $F_{\rm r}(x,v)$ and $F_{\rm p}(x,v)$, as a function of energy reserves $x$ and enamel volume $v$ for both non-megadont and megadont hominins when grass leaves are hyper-abundant and for the arthropod + USO mechanical advantage scenario. 
      Gray elements to the left and bottom of the plots denote values of $(x,v)$ resulting in mortality.
      }
      \label{fig:AbLeaves}
\end{figure*}

Under the conditions imposed by our model, ${\rm C_4}$ grass leaves cannot maximize fitness.
However, we can explore under what conditions grass leaves do maximize fitness by altering model properties.
We find that grass leaves become represented in the decision matrices of hominins both with and without megadontia if the abundance of grass is exaggerated (such that the encounter rate of grass leaves is increased from 4 to 5; Figure \ref{fig:AbLeaves}A,B).
Even then, the consumption of grass leaves is shown to be a fallback behavior {\it in extremis}, selected only when enamel volume is high and energy reserves are extremely low. 
Moreover, megadontia leads to a relatively greater percentage of states where grass leaves maximize fitness (Figure \ref{fig:AbLeaves}B), and this is in accordance with the elevated $\delta^{13}{\rm C}$ values observed for species in the genus {\it Paranthropus}.

\section*{\Large  \sf{\emph{Discussion}}}

Models have been used to explore the foraging behaviors of humans \citep{Belovsky:1988p3438}, nonhuman primates \citep{Boyer:2006gz,Sayers:2010ba}, and their mutual interactions \citep{Levi:2011tk}, but few have been applied to extinct primates \citep{Dunbar:2005ug,Janssen:2007vf,Griffith:2010uq}, and none have accounted for nonrenewable resources such as dental enamel.
This omission is surprising given the functional and adaptive significance prescribed to molar enamel thickness.
In this vein, a Stochastic Dynamic Programming (SDP) model is attractive because it demands the explicit expression of processes that determine fitness, as well as sources of external and internal stochasticity \citep{Mangel:1988uaa,Clark:2000tra}.
We have developed an SDP model that assesses directly the role of enamel volume on food selection and fitness while quantifying the extent to which anatomical and behavioral attributes can alter foraging behaviors.

\subsection*{\small \sf{\textbf{THICK ENAMEL CONFERS A FITNESS ADVANTAGE}}}

\begin{figure*}[h]
   \centering
   \includegraphics[width=1\textwidth]{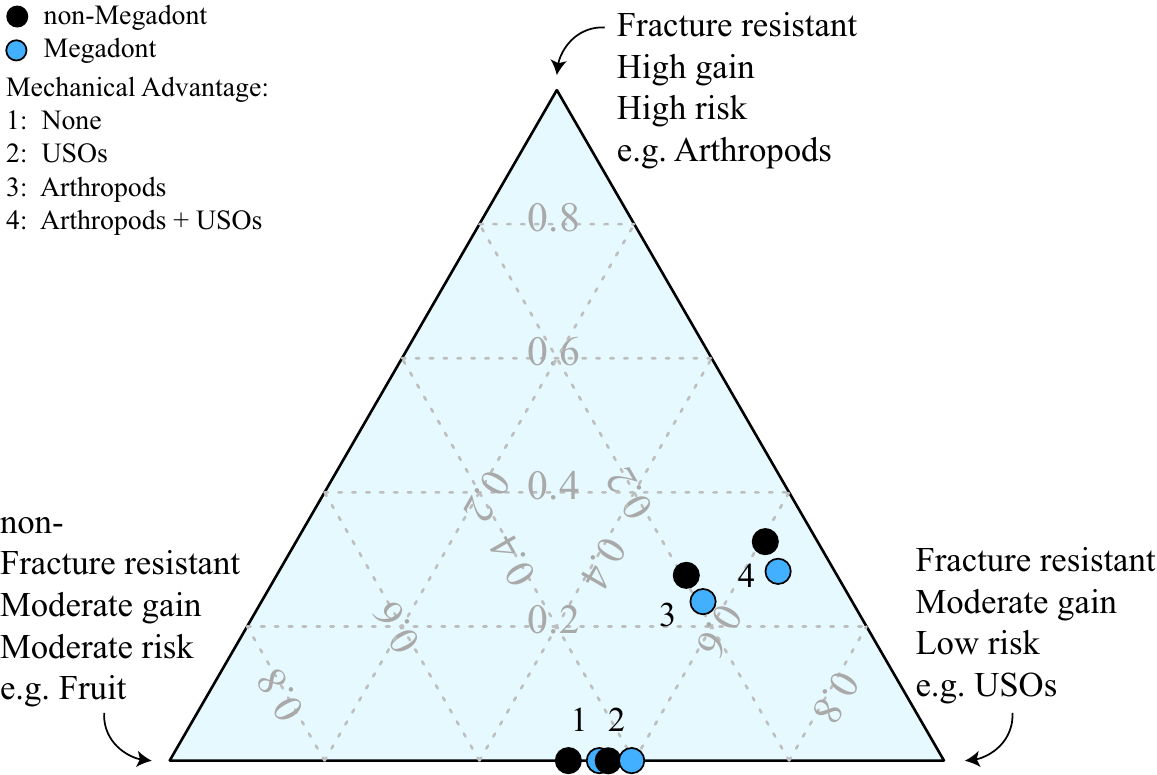}
      \caption{
      Ternary diagram showing the proportional contribution of fruit, USOs, and arthropods to the decision matrices of both 50 kg non-megadont and megadont primates under each mechanical advantage scenario. 
      Results are shown for autocorrelated environmental conditions; results for wet and dry conditions were qualitatively similar.
      }
      \label{fig:Tern}
\end{figure*}

The relatively massive molar teeth of {\it Paranthropus} are invested with hyperthick enamel \citep{Shellis:1998fp,Lucas:2008p1237}.
This combination of traits, or megadontia, is coupled with robust jaws and large chewing muscles, which together enable an immense bite force \citep{Demes:1988ct,Constantino:2010et}.
Functional interpretations of these traits have long stressed the consumption of hard or obdurate foods \citep{Kay:1981tr,Osborne:wx,Macho:1999tc}, although a recent trend has emphasized tough foods that require repetitive loading (grinding) of the jaws and teeth \citep{Ungar:2011em}, particularly with respect to {\it P. boisei} \citep{Ungar:2008tj,Ungar:2012bp}.
In either case, debate has focused on a diet of grass seeds \citep{Jolly:1970wf} or plant underground storage organs (USOs) as the primary drivers of this robust morphology \citep{Laden:2005p44,Sponheimer:2005p1377,Yeakel:2007p1410,Dominy:2008p864}.
The results of our SDP model agree well with these hypotheses by showing that hyperthick molar enamel reduces the mechanical costs of chewing harder foods over greater proportions of internal states $(x,v)$  (Figure \ref{fig:Tern}).
Megadontia, then, provides an adaptive advantage in poor quality environments where hard foods such as grass seeds and USOs are relatively abundant.

Hominins were doubtless tool-users, and the ability to alter the physical properties of wear-inducing foods is expected to both increase dietary breadth and decrease fitness costs.
In support of this prediction, the inclusion of an extradentary mechanical advantage in our model increased the proportion of high-risk foods in the predicted decision matrices (Figure \ref{fig:Tern}). 
A USO mechanical advantage increased the proportion of USOs in the diet, albeit marginally, whereas the consumption of fruit declined.
By comparison, the extradentary mechanical advantage associated with arthropods or both arthropods and USOs had a large effect on the decision matrices. 
Arthropods were fitness-maximizing foods for hominins both with and without megadontia because they decreased the risk of obtaining rare or patchily distributed foods, while reducing their reliance on fruit. 
Extradentary processing is therefore advantageous; however, it is telling that USOs always maximized fitness across a greater proportion of states for hominins with greater enamel volume.

Importantly, the predicted fitness advantages of thick enamel are variable due to the different rates of enamel wear over a lifetime (Figure \ref{fig:fit}). 
In this regard, our process-based model is relatively simplistic in that life-history stages are excluded; however, these simplifications enabled us to test and affirm three predictions regarding hominin foraging behavior:
1) behaviors that alter the mechanical properties of hard foods result in greater fitness; 
2) these benefits are primarily realized in dry environments, where habitat quality is more likely to be poor and hard foods are relatively more abundant; and,
3) because megadontia results in relatively slower rates of wear, it confers relatively higher fitness, and these benefits are primarily realized later in life.

In summary, our SDP model demonstrates that different foraging choices are predicted to maximize fitness among hominins with varying degrees of megadontia, and that these foraging strategies have different expected lifetime fitness values.
In the following sections we discuss how a forward iteration approach can be used to examine the isotopic differences observed among hominin species, and whether the mechanical and physiological constraints imposed by our model are predictive of the isotopic patterns observed in the fossil record.

\subsection*{\small \sf{\textbf{COMPARING MODEL PREDICTIONS TO ISOTOPIC DATA}}}

Results from simulations of the $\delta^{13}{\rm C}$ values accumulated over a lifetime of a hominin population help to resolve occasional discrepancies between craniodental morphology (indicating hard foods) and molar microwear (indicating soft foods) \citep{Grine:2012gw}.
Molar enamel is formed early in life \citep{Lucas:2004wz} when food selection tends towards mechanically hard foods with high $\delta^{13}{\rm C}$ values (Figure \ref{fig:IsoSim}A,B).
As enamel is worn, softer, less abundant foods with lower $\delta^{13}{\rm C}$ values are shown to maximize fitness.
Because fossilized microwear is formed shortly before death (the `last supper effect'), our model results suggest that softer, more pliable foods will have a disproportionately large influence on the microwear of teeth, particularly for older individuals.
Moreover, simulated foragers incorporated foods in proportions that are not predicted by their relative abundance on the landscape (Figure \ref{fig:IsoSim}B,D), highlighting the importance of considering both mechanical and energetic constraints in addition to resource abundance. 

\subsection*{\small\sf{\textbf{FALLBACK FOODS ARE BODY SIZE-DEPENDENT}}}

Multiple lines of evidence suggest that plant USOs were important foods for early hominins.
Plants with geophytic structures are both diverse and abundant in arid habitats \citep{Pate:1982vt,Vincent:1985hn,Proches:2006ea}, and modern hunter-gatherers utilize these resources extensively, particularly in marginal environments \citep{Campbell:1986vs,Marlowe:2003p1049,Marlowe:2009p1554}.
Associations between mole rats - known USO specialists - and hominins suggest that human ancestors lived in USO-abundant habitats \citep{Laden:2005p44}, and stable isotope analysis of both modern and fossil mole rats confirm that USO specialists have isotopic values similar to those of {\it A. africanus} and {\it P. robustus} \citep{Yeakel:2007p1410}. 
It is widely assumed that USOs served as fallback rather than preferred foods due to their lower nutritional content and relative availability \citep{Schoeninger:2001ht}.
The results of our model are in general agreement with this assumption, but show that the role of USOs as fallback foods varies - in part - as a function of an organism's energy reserves and enamel volume, as well as body size.

In general, the consumption of USOs is predicted if enamel volume is relatively high and energy reserves are relatively low (Figure \ref{fig:DM_mn}).
However, our model also predicts a tradeoff with respect to the role of USOs as fallback foods as body size is altered.
Smaller-sized animals tend to use nutritious foods such as fruit in rich quality habitats and less nutritious but more ubiquitous foods such as USOs in poor quality habitats (Figure \ref{fig:PercFood}A).
Thus, as energetic reserves become more limiting, as they are for smaller organisms with relatively higher resting metabolic rates, fruit and USOs alternatively serve as preferred foods when habitat quality is rich and poor, respectively.
By comparison, larger body size enables risky foraging even when habitat quality is poor, and such risky foraging becomes commonplace if an organism can mechanically alter its food (Figure \ref{fig:PercFood}B,C,D). 
For all scenarios, larger animals resort to USO consumption when energy reserves are low.
Accordingly, USOs are relegated to a fallback status, and are consumed if the act of foraging for preferred foods incurs relatively greater fitness costs on the organism.
Although consumption of USOs reduces the costs of foraging in poor quality environments, our results also show that widespread but nutritionally poor and mechanically obdurate foods such as grass leaves are actively avoided, even when there are enamel and energetic reserves to spare.

\subsection*{\small \sf{\textbf{GRASS LEAVES DO NOT MAXIMIZE FITNESS}}}

Despite the ubiquity of ${\rm C_4}$ grass leaves in hominin habitats, this potential food resource is an unlikely solution to the SDP, consistent with the aversion to ${\rm C_4}$ plants that is evident among savanna-dwelling chimpanzees \citep{Sponheimer:2006p718}, modern lemurs \citep{Crowley:2013to}, and some hominin species including {\it Ardipithecus ramidus} \citep{White:2009p2395} and {\it Australopithecus sediba} \citep{Henry:2012bj}.
Because we used a conservatively low value for the fracture toughness of ${\rm C}_4$ grass leaves (see methods), the absence of this food from hominin decision matrices is a telling argument against the concept of a grazing hominin.
The underlying reasons for this aversion are unknown, but ${\rm C_4}$ grass leaves are often more fracture-resistant \citep{Boutton:1978ti} and less nutritious \citep{Barbehenn:2004p734} than ${\rm C_3}$ leaves, possibly due to the presence of bundle sheath cells.
These factors have been cited to explain the avoidance of ${\rm C_4}$ plants by herbaceous insects in grassland communities \citep{Caswell:1973p1272,Boutton:1978ti,PinderIII:1987p1266}.

Yet, megadont hominins such as {\it P. boisei} have $\delta^{13}{\rm C}$ values $\approx 0 \permil$, which corresponds to a diet of 75-80\% $\rm C_4$ foods \citep{Ungar:2011em}.
Such a heavy dependence on ${\rm C}_4$ foods has led to speculation that {\it P. boisei} was potentially a grazing hominin \citep{LeeThorp:2011ia,Rabenold:2011gw}. 
Our model results indicate that grass leaves do have the potential to maximize fitness in extreme circumstances, though the benefits of this food source decline quickly as enamel is worn. 
This suggests that ${\rm C}_4$ grass leaves are unlikely to confer fitness advantages even for hominins with megadontia.


\subsection*{\small \sf{\textbf{CONCLUSION}}}

Foraging behaviors are a consequence of both the mechanical and energetic costs of food and the constraints imposed by an organism's dentition.
Dental enamel thickness is a highly conserved trait among individuals within modern human populations \citep{Lucas:2008bd}, yet it varies considerably across hominin lineages in the fossil record.
This variability is an evolutionary consequence of interactions between the dentition and food, and process-based models that integrate these ingredients can inform both the possible roles of certain foods as well as the potential fitness benefits of different dental morphologies or extradentary mechanical advantages.
Along this line, a similar SDP approach could be used to investigate the roles of different types of USOs - foods that include corms, tubers, bulbs, and rhizomes.
Because these plant parts are distributed differently across ${\rm C}_3$ and ${\rm C}_4$ plant species, preference or avoidance of such potential foods - as a function of energetic and mechanical gains and costs - may help explain the surprisingly high $\delta^{13}{\rm C}$ values of hominins such as {\it P. boisei}. 
Regardless, we believe that the integration of data obtained from the fossil record with mechanistic models that set physical constraints on potential behaviors will expand our understanding of these enigmatic organisms.

\acknowledgements{{\bf Acknowledgements:} We thank CE Chow, AM Kilpatrick, TS Kraft, T Levi, PW Lucas, AD Melin, GL Moritz, M Novak, AO Shelton, and ER Vogel for insightful comments and helpful discussions. This work was partially supported by a National Science Foundation (NSF) Grant 2009-0417 to MM, and a NSF-GRF to JDY. The authors declare no conflicts of interest.}

\newpage


\begin{thebibliography}{92}
\providecommand{\natexlab}[1]{#1}
\providecommand{\url}[1]{\texttt{#1}}
\providecommand{\urlprefix}{URL }

\bibitem[{Alberts et~al.(2005)Alberts, Hollister-Smith, Mututua, Sayialel,
  Muruthi, Warutere, and Altmann}]{Alberts:2005ug}
Alberts, S.~C., J.~A. Hollister-Smith, R.~S. Mututua, S.~N. Sayialel, P.~M.
  Muruthi, J.~K. Warutere, and J.~Altmann, 2005.
\newblock {Seasonality and long-term change in a savanna environment}.
\newblock \emph{in} D.~K. Brockman and C.~P. van Schaik, eds. Seasonality in
  primates: Studies of living and extinct human and non-human primates.
  Cambridge University Press, Cambridge.

\bibitem[{Altmann(2009)}]{Altmann:2009el}
Altmann, S.~A., 2009.
\newblock {Fallback foods, eclectic omnivores, and the packaging problem}.
\newblock Am. J. Phys. Anthropol. 140:615--629.

\bibitem[{Balter et~al.(2012)Balter, Braga, T{\'e}louk, and
  Thackeray}]{Balter:2012wt}
Balter, V., J.~Braga, P.~T{\'e}louk, and J.~F. Thackeray, 2012.
\newblock {Evidence for dietary change but not landscape use in South African
  early hominins}.
\newblock Nature 489:558--560.

\bibitem[{Barbehenn et~al.(2004)Barbehenn, Chen, Karowe, and
  Spickard}]{Barbehenn:2004p734}
Barbehenn, R., Z.~Chen, D.~Karowe, and A.~Spickard, 2004.
\newblock {${\rm C_3}$ grasses have higher nutritional quality than ${\rm C_4}$ grasses under
  ambient and elevated atmospheric $\rm CO_2$}.
\newblock Global Change Biol. 10:1565--1575.

\bibitem[{Barton et~al.(1993)Barton, Whiten, Byrne, and
  English}]{Barton:1993wc}
Barton, R.~A., A.~Whiten, R.~W. Byrne, and M.~English, 1993.
\newblock {Chemical composition of baboon plant foods: implications for the
  interpretation of intra- and interspecific differences in diet.}
\newblock Folia Primatol. 61:1--20.

\bibitem[{Belovsky(1988)}]{Belovsky:1988p3438}
Belovsky, G., 1988.
\newblock {An optimal foraging-based model of hunter-gatherer population
  dynamics}.
\newblock J. Anthropol. Archaeol. 7:329--372.

\bibitem[{Boccara(2004)}]{Boccara:2004p2506}
Boccara, N., 2004.
\newblock {Modeling complex systems‎}.
\newblock Springer, New York.

\bibitem[{Boutton et~al.(1978)Boutton, Cameron, and Smith}]{Boutton:1978ti}
Boutton, T., G.~Cameron, and B.~Smith, 1978.
\newblock {Insect herbivory on ${\rm C_3}$ and ${\rm C_4}$ grasses}.
\newblock Oecologia 36:21--32.

\bibitem[{Boyer et~al.(2006)Boyer, Ramos-Fernandez, Miramontes, Mateos, Cocho,
  Larralde, Ramos, and Rojas}]{Boyer:2006gz}
Boyer, D., G.~Ramos-Fernandez, O.~Miramontes, J.~L. Mateos, G.~Cocho,
  H.~Larralde, H.~Ramos, and F.~Rojas, 2006.
\newblock {Scale-free foraging by primates emerges from their interaction with
  a complex environment}.
\newblock Proc. Roy. Soc. B 273:1743--1750.

\bibitem[{Bronikowski et~al.(2011)Bronikowski, Altmann, Brockman, Cords,
  Fedigan, Pusey, Stoinski, Morris, Strier, and Alberts}]{Bronikowski:2011jd}
Bronikowski, A.~M., J.~Altmann, D.~K. Brockman, M.~Cords, L.~M. Fedigan,
  A.~Pusey, T.~Stoinski, W.~F. Morris, K.~B. Strier, and S.~C. Alberts, 2011.
\newblock {Aging in the natural world: comparative data reveal similar
  mortality patterns across primates}.
\newblock Science 331:1325--1328.

\bibitem[{Campbell(1986)}]{Campbell:1986vs}
Campbell, A., 1986.
\newblock {The use of wild food plants, and drought in Botswana}.
\newblock J. Arid Environ. 11:81--91.

\bibitem[{Caswell et~al.(1973)Caswell, Reed, Stephenson, and
  Werner}]{Caswell:1973p1272}
Caswell, H., F.~Reed, S.~N. Stephenson, and P.~Werner, 1973.
\newblock {Photosynthetic pathways and selective herbivory: a hypothesis}.
\newblock Am. Nat. 107:465--480.

\bibitem[{Cerling et~al.(2011)Cerling, Mbua, Kirera, Manthi, Grine, Leakey,
  Sponheimer, and Uno}]{Cerling:2011wh}
Cerling, T.~E., E.~Mbua, F.~M. Kirera, F.~K. Manthi, F.~E. Grine, M.~G. Leakey,
  M.~Sponheimer, and K.~T. Uno, 2011.
\newblock {Diet of {\it Paranthropus boisei} in the early Pleistocene of East
  Africa}.
\newblock Proc. Natl. Acad. Sci. USA 108:9337--9341.

\bibitem[{Clark and Mangel(2000)}]{Clark:2000tra}
Clark, C.~W. and M.~Mangel, 2000.
\newblock {Dynamic state variable models in ecology: methods and applications}.
\newblock Oxford University Press, USA.

\bibitem[{Constantino(2013)}]{Constantino:2013ut}
Constantino, P.~J., 2013.
\newblock {The ``Robust'' Australopiths}.
\newblock Nature Educ. Knowl. 4:1.

\bibitem[{Constantino et~al.(2010)Constantino, Lee, Chai, Zipfel, Ziscovici,
  Lawn, and Lucas}]{Constantino:2010et}
Constantino, P.~J., J.~J.~W. Lee, H.~Chai, B.~Zipfel, C.~Ziscovici, B.~R. Lawn,
  and P.~W. Lucas, 2010.
\newblock {Tooth chipping can reveal the diet and bite forces of fossil
  hominins}.
\newblock Biol. Letters 6:826--829.

\bibitem[{Copeland(2004)}]{Copeland:2004to}
Copeland, S., 2004.
\newblock {Paleoanthropological implications of vegetation and wild plant
  resources in modern savanna landscapes, with applications to Plio-Pleistocene
  Olduvai Gorge, Tanzania}.
\newblock Ph.D. thesis, Rutgers University.

\bibitem[{Crowley and Samonds(2013)}]{Crowley:2013to}
Crowley, B.~E. and K.~E. Samonds, 2013.
\newblock {Stable carbon isotope values confirm a recent increase in grasslands
  in northwestern Madagascar}.
\newblock Holocene 23:1066--1073.

\bibitem[{Demes and Creel(1988)}]{Demes:1988ct}
Demes, B. and N.~Creel, 1988.
\newblock {Bite force, diet, and cranial morphology of fossil hominids}.
\newblock J. Hum. Evol. 17:657--670.

\bibitem[{Dominy(2012)}]{Dominy:2012bi}
Dominy, N.~J., 2012.
\newblock {Hominins living on the sedge}.
\newblock Proc. Natl. Acad. Sci. USA 109:20171--20172.

\bibitem[{Dominy et~al.(2008)Dominy, Vogel, Yeakel, Constantino, and
  Lucas}]{Dominy:2008p864}
Dominy, N.~J., E.~R. Vogel, J.~D. Yeakel, P.~J. Constantino, and P.~W. Lucas,
  2008.
\newblock {Mechanical properties of plant underground storage organs and
  implications for dietary models of early hominins}.
\newblock Evol. Biol. 35:159--175.

\bibitem[{Dunbar(2005)}]{Dunbar:2005ug}
Dunbar, R. I.~M., 2005.
\newblock {Socioecology of the extinct theropiths: a modelling approach}.
\newblock \emph{in} {\it Theropithecus}: the rise and fall of a primate genus.
  Cambridge University Press, Cambridge.

\bibitem[{Galbany et~al.(2011)Galbany, Altmann, P{\'e}rez-P{\'e}rez, and
  Alberts}]{Galbany:2011cm}
Galbany, J., J.~Altmann, A.~P{\'e}rez-P{\'e}rez, and S.~C. Alberts, 2011.
\newblock {Age and individual foraging behavior predict tooth wear in Amboseli
  baboons.}
\newblock Am. J. Phys. Anthropol. 144:51--59.

\bibitem[{Griffith and Long(2010)}]{Griffith:2010uq}
Griffith, C. and B.~Long, 2010.
\newblock {HOMINIDS: An agent-based spatial simulation model to evaluate
  behavioral patterns of early Pleistocene hominids}.
\newblock Ecol. Model. 221:738--760.

\bibitem[{Grine et~al.(2012)Grine, Sponheimer, Ungar, Lee-Thorp, and
  Teaford}]{Grine:2012gw}
Grine, F.~E., M.~Sponheimer, P.~S. Ungar, J.~Lee-Thorp, and M.~F. Teaford,
  2012.
\newblock {Dental microwear and stable isotopes inform the paleoecology of
  extinct hominins.}
\newblock Am. J. Phys. Anthropol. 148:285--317.

\bibitem[{Henry et~al.(2012)Henry, Ungar, Passey, Sponheimer, Rossouw, Bamford,
  Sandberg, de~Ruiter, and Berger}]{Henry:2012bj}
Henry, A.~G., P.~S. Ungar, B.~H. Passey, M.~Sponheimer, L.~Rossouw, M.~Bamford,
  P.~Sandberg, D.~J. de~Ruiter, and L.~Berger, 2012.
\newblock {The diet of {\it Australopithecus sediba}}.
\newblock Nature 487:90--93.

\bibitem[{Houston and McNamara(1999)}]{Houston:1999va}
Houston, A. and J.~M. McNamara, 1999.
\newblock {Models of adaptive behaviour}.
\newblock Cambridge University Press, Cambridge.

\bibitem[{Janssen et~al.(2007)Janssen, Sept, and Griffith}]{Janssen:2007vf}
Janssen, M., J.~Sept, and C.~Griffith, 2007.
\newblock {Hominids foraging in a complex landscape: Could {\it Homo ergaster}
  and {\it Australopithecus boisei} meet their calories requirements?}
\newblock Takahashi, S., Sallach, D., {\&} and Rouchier, J.(Eds.), Advancing
  Social Simulation. Springer Publishing Pp. 307--318.

\bibitem[{Jolly(1970)}]{Jolly:1970wf}
Jolly, C.~J., 1970.
\newblock {The seed-eaters: a new model of hominid differentiation based on a
  baboon analogy}.
\newblock Man 5:1--26.

\bibitem[{Juanes(1992)}]{Juanes:1992uv}
Juanes, F., 1992.
\newblock {Why do decapod crustaceans prefer small-sized molluscan prey?}
\newblock Mar. Ecol.-Prog. Ser. 87:239--239.

\bibitem[{Juanes and Smith(1995)}]{Juanes:1995jx}
Juanes, F. and L.~Smith, 1995.
\newblock {The ecological consequences of limb damage and loss in decapod
  crustaceans: a review and prospectus}.
\newblock J. Exp. Mar. Biol. Ecol. 193:197--223.

\bibitem[{Kay(1981)}]{Kay:1981tr}
Kay, R.~F., 1981.
\newblock {The nut-crackers: A new theory of the adaptations of the
  Ramapithecinae}.
\newblock Am. J. Phys. Anthropol. 55:141--151.

\bibitem[{Kay(1985)}]{Kay:1985p987}
---{}---{}---, 1985.
\newblock {Dental evidence for the diet of {\it Australopithecus}}.
\newblock Annu. Rev. Anthropol. 14:315--341.

\bibitem[{King et~al.(2005)King, Arrigo-Nelson, Pochron, Semprebon, Godfrey,
  Wright, and Jernvall}]{King:2005kj}
King, S.~J., S.~J. Arrigo-Nelson, S.~T. Pochron, G.~M. Semprebon, L.~R.
  Godfrey, P.~C. Wright, and J.~Jernvall, 2005.
\newblock {Dental senescence in a long-lived primate links infant survival to
  rainfall}.
\newblock Proc. Natl. Acad. Sci. USA 102:16579--16583.

\bibitem[{Laden and Wrangham(2005)}]{Laden:2005p44}
Laden, G. and R.~Wrangham, 2005.
\newblock {The rise of the hominids as an adaptive shift in fallback foods:
  plant underground storage organs (USOs) and australopith origins.}
\newblock J. Hum. Evol. 49:482--498.

\bibitem[{Lee-Thorp(2011)}]{LeeThorp:2011ia}
Lee-Thorp, J., 2011.
\newblock {The demise of "Nutcracker Man".}
\newblock Proc. Natl. Acad. Sci. USA 108:9319--9320.

\bibitem[{Lee-Thorp et~al.(2012)Lee-Thorp, Likius, Mackaye, Vignaud,
  Sponheimer, and Brunet}]{LeeThorp:2012ea}
Lee-Thorp, J., A.~Likius, H.~T. Mackaye, P.~Vignaud, M.~Sponheimer, and
  M.~Brunet, 2012.
\newblock {Isotopic evidence for an early shift to ${\rm C_4}$ resources by Pliocene
  hominins in Chad.}
\newblock Proc. Natl. Acad. Sci. USA 109:20369--20372.

\bibitem[{Leonard and Robertson(1997)}]{Leonard:1997usa}
Leonard, W. and M.~Robertson, 1997.
\newblock {Comparative primate energetics and hominid evolution}.
\newblock Am. J. Phys. Anthropol. 102:265--281.

\bibitem[{Levi et~al.(2011)Levi, Lu, Yu, and Mangel}]{Levi:2011tk}
Levi, T., F.~Lu, D.~Yu, and M.~Mangel, 2011.
\newblock {The behaviour and diet breadth of central-place foragers: an
  application to human hunters and Neotropical game management}.
\newblock Evol. Ecol. Res. 13:171--185.

\bibitem[{Lindstedt and Calder~III(1981)}]{Lindstedt:1981p1038}
Lindstedt, S.~L. and W.~A. Calder~III, 1981.
\newblock {Body size, physiological time, and longevity of homeothermic
  animals}.
\newblock Q. Rev. Biol. 56:1--16.

\bibitem[{Lucas(2004)}]{Lucas:2004wz}
Lucas, P.~W., 2004.
\newblock {Dental functional morphology: How teeth work}.
\newblock Cambridge University Press, Cambridge.

\bibitem[{Lucas et~al.(2008{\natexlab{a}})Lucas, Constantino, and
  Wood}]{Lucas:2008p1237}
Lucas, P.~W., P.~J. Constantino, and B.~A. Wood, 2008{\natexlab{a}}.
\newblock {Inferences regarding the diet of extinct hominins: structural and
  functional trends in dental and mandibular morphology within the hominin
  clade}.
\newblock J. Anat. 212:486--500.

\bibitem[{Lucas et~al.(2008{\natexlab{b}})Lucas, Constantino, Wood, and
  Lawn}]{Lucas:2008bd}
Lucas, P.~W., P.~J. Constantino, B.~A. Wood, and B.~R. Lawn,
  2008{\natexlab{b}}.
\newblock {Dental enamel as a dietary indicator in mammals}.
\newblock Bioessays 30:374--385.

\bibitem[{Lucas et~al.(2012)Lucas, Omar, Al-Fadhalah, Almusallam, Henry,
  Michael, Thai, Watzke, Strait, and Atkins}]{Lucas:2012gj}
Lucas, P.~W., R.~Omar, K.~Al-Fadhalah, A.~S. Almusallam, A.~G. Henry,
  S.~Michael, L.~A. Thai, J.~Watzke, D.~S. Strait, and A.~G. Atkins, 2012.
\newblock {Mechanisms and causes of wear in tooth enamel: implications for
  hominin diets}.
\newblock J. R. Soc. Interface 10:20120923.

\bibitem[{Lucas et~al.(2000)Lucas, Turner, Dominy, and
  Yamashita}]{Lucas:2000bca}
Lucas, P.~W., I.~M. Turner, N.~J. Dominy, and N.~Yamashita, 2000.
\newblock {Mechanical defences to herbivory}.
\newblock Ann Bot-London 86:913--920.

\bibitem[{Macho(1999)}]{Macho:1999tc}
Macho, G.~A., 1999.
\newblock {Effects of loading on the biochemical behavior of molars of {\it
  Homo}, {\it Pan}, and {\it Pongo}}.
\newblock Am. J. Phys. Anthropol. 109:211--227.

\bibitem[{Mangel and Clark(1988)}]{Mangel:1988uaa}
Mangel, M. and C.~W. Clark, 1988.
\newblock {Dynamic modeling in behavioral ecology}.
\newblock Princeton University Press, Princeton.

\bibitem[{Mangel and Ludwig(1992)}]{Mangel:1992wo}
Mangel, M. and D.~Ludwig, 1992.
\newblock {Definition and evaluation of the fitness of behavioral and
  developmental programs}.
\newblock Annu. Rev. Ecol. Syst. 23:507--536.

\bibitem[{Marlowe(2003)}]{Marlowe:2003p1049}
Marlowe, F.~W., 2003.
\newblock {A critical period for provisioning by Hadza men implications for
  pair bonding}.
\newblock Evol. Hum. Behav. 24:217--229.

\bibitem[{Marlowe and Berbesque(2009)}]{Marlowe:2009p1554}
Marlowe, F.~W. and J.~C. Berbesque, 2009.
\newblock {Tubers as fallback foods and their impact on Hadza
  hunter-gatherers}.
\newblock Am. J. Phys. Anthropol. 140:751--758.

\bibitem[{McArthur and Sanson(1988)}]{McArthur:1988kd}
McArthur, C. and G.~D. Sanson, 1988.
\newblock {Tooth wear in eastern grey kangaroos ({\it Macropus giganteus}) and
  western grey kangaroos ({\it Macropus fuliginosus}), and its potential
  influence on diet selection, digestion and population parameters}.
\newblock J. Zool. 215:491--504.

\bibitem[{McGraw and Daegling(2012)}]{McGraw:2012dj}
McGraw, W.~S. and D.~J. Daegling, 2012.
\newblock {Primate Feeding and Foraging: Integrating Studies of Behavior and
  Morphology}.
\newblock Annu. Rev. Anthropol. 41:203--219.

\bibitem[{McGraw et~al.(2012)McGraw, Pampush, and Daegling}]{McGraw:2012fl}
McGraw, W.~S., J.~D. Pampush, and D.~J. Daegling, 2012.
\newblock {Brief communication: Enamel thickness and durophagy in mangabeys
  revisited.}
\newblock Am. J. Phys. Anthropol. 147:326--333.

\bibitem[{McHenry(1988)}]{McHenry:1988tb}
McHenry, H.~M., 1988.
\newblock {New estimates of body weights in early hominids and their
  significance to encephalization and megadontia in ``robust''
  australopithecines}.
\newblock \emph{in} Evolutionary History of the "Robust" Australopithecines.
  Transaction Publishers, New Brunswick.

\bibitem[{Milton(1989)}]{Milton:1989vt}
Milton, K., 1989.
\newblock {Primate diets and gut morphology: implications for hominid
  evolution}.
\newblock \emph{in} Food and evolution: toward a theory of human food habits.
  Temple University Press, Philadelphia.

\bibitem[{Organ et~al.(2011)Organ, Nunn, Machanda, and Wrangham}]{Organ:2011td}
Organ, C., C.~L. Nunn, Z.~Machanda, and R.~W. Wrangham, 2011.
\newblock {Phylogenetic rate shifts in feeding time during the evolution of
  {\it Homo}}.
\newblock Proc. Natl. Acad. Sci. USA 108:14555--14559.

\bibitem[{Osborne(1981)}]{Osborne:wx}
Osborne, J.~W., 1981.
\newblock {Ageing}.
\newblock Pp. 352--356, \emph{in} A.~H.~R. Rowe and R.~B. Johns, eds. Dental
  Anatomy and Embryology Vol. 2: A comparison to dental studies. Blackwell,
  Oxford.

\bibitem[{Pate and Dixon(1982)}]{Pate:1982vt}
Pate, J.~S. and K.~W. Dixon, 1982.
\newblock {Tuberous, cormous and bulbous plants: biology of an adaptive
  strategy in Western Australia}.
\newblock University of Western Australia Press, Nedlands, Australia.

\bibitem[{Perez-Barberia and Gordon(1998)}]{PerezBarberia:1998gj}
Perez-Barberia, F.~J. and I.~J. Gordon, 1998.
\newblock {The influence of molar occlusal surface area on the voluntary
  intake, digestion, chewing behaviour and diet selection of red deer ({\it
  Cervus elaphus})}.
\newblock J. Zool. 245:307--316.

\bibitem[{Pinder~III and Kroh(1987)}]{PinderIII:1987p1266}
Pinder~III, J.~E. and G.~C. Kroh, 1987.
\newblock {Insect herbivory and photosynthetic pathways in old-field
  ecosystems}.
\newblock Ecology 68:254--259.

\bibitem[{Post(1982)}]{Post:1982bt}
Post, D.~G., 1982.
\newblock {Feeding behavior of yellow baboons ({\it Papio cynocephalusin}) the
  Amboseli National Park, Kenya}.
\newblock Int. J. Primatol. 3:403--430.

\bibitem[{Proche{\c s} et~al.(2006)Proche{\c s}, Cowling, Goldblatt, Manning,
  and Snijman}]{Proches:2006ea}
Proche{\c s}, {\c S}., R.~M. Cowling, P.~Goldblatt, J.~C. Manning, and D.~A.
  Snijman, 2006.
\newblock {An overview of the Cape geophytes}.
\newblock Biol. J. Linn. Soc. 87:27--43.

\bibitem[{Rabenold and Pearson(2011)}]{Rabenold:2011gw}
Rabenold, D. and O.~M. Pearson, 2011.
\newblock {Abrasive, silica phytoliths and the evolution of thick molar enamel
  in primates, with implications for the diet of {\it Paranthropus boisei}}.
\newblock PLoS ONE 6:e28379.

\bibitem[{Raupp(1985)}]{Raupp:1985ub}
Raupp, M.~J., 1985.
\newblock {Effects of leaf toughness on mandibular wear of the leaf beetle,
  {\it Plagiodera versicolora}}.
\newblock Ecol. Entomol. 10:73--79.

\bibitem[{Roitberg et~al.(2005)Roitberg, Gillespie, Quiring, Alma, Jenner,
  Perry, Peterson, Salomon, and VanLaerhoven}]{Roitberg:2005iv}
Roitberg, B.~D., D.~R. Gillespie, D.~M.~J. Quiring, C.~R. Alma, W.~H. Jenner,
  J.~Perry, J.~H. Peterson, M.~Salomon, and S.~VanLaerhoven, 2005.
\newblock {The cost of being an omnivore: mandible wear from plant feeding in a
  true bug.}
\newblock Naturwissenschaften 92:431--434.

\bibitem[{Rothman et~al.(2011)Rothman, Raubenheimer, and
  Chapman}]{Rothman:2011kh}
Rothman, J.~M., D.~Raubenheimer, and C.~A. Chapman, 2011.
\newblock {Nutritional geometry: gorillas prioritize non-protein energy while
  consuming surplus protein.}
\newblock Biol. Letters 7:847--849.

\bibitem[{Sayers et~al.(2010)Sayers, Norconk, and
  Conklin-Brittain}]{Sayers:2010ba}
Sayers, K., M.~A. Norconk, and N.~L. Conklin-Brittain, 2010.
\newblock {Optimal foraging on the roof of the world: Himalayan langurs and the
  classical prey model.}
\newblock Am. J. Phys. Anthropol. 141:337--357.

\bibitem[{Schoeninger et~al.(2001)Schoeninger, Bunn, Murray, and
  Marlett}]{Schoeninger:2001ht}
Schoeninger, M.~J., H.~T. Bunn, S.~S. Murray, and J.~A. Marlett, 2001.
\newblock {Composition of tubers used by Hadza foragers of Tanzania}.
\newblock J. Food Compos. Anal. 14:15--25.

\bibitem[{Shellis et~al.(1998)Shellis, Beynon, Reid, and
  Hiiemae}]{Shellis:1998fp}
Shellis, R.~P., A.~D. Beynon, D.~J. Reid, and K.~M. Hiiemae, 1998.
\newblock {Variations in molar enamel thickness among primates.}
\newblock J. Hum. Evol. 35:507--522.

\bibitem[{Skogland(1988)}]{Skogland:1988wt}
Skogland, T., 1988.
\newblock {Tooth wear by food limitation and its life history consequences in
  wild reindeer}.
\newblock Oikos 51:238--242.

\bibitem[{Sponheimer and Lee-Thorp(1999)}]{Sponheimer:1999p1389}
Sponheimer, M. and J.~Lee-Thorp, 1999.
\newblock {Isotopic evidence for the diet of an early hominid, {\it
  Australopithecus africanus}}.
\newblock Science 283:368.

\bibitem[{Sponheimer and Lee-Thorp(2003)}]{Sponheimer:2003p1378}
---{}---{}---, 2003.
\newblock {Differential resource utilization by extant great apes and
  australopithecines: towards solving the ${\rm C_4}$ conundrum}.
\newblock Comp. Biochem. Phys. A 136:27--34.

\bibitem[{Sponheimer et~al.(2005)Sponheimer, Lee-Thorp, de~Ruiter, Codron,
  Codron, Baugh, and Thackeray}]{Sponheimer:2005p1377}
Sponheimer, M., J.~Lee-Thorp, D.~de~Ruiter, D.~Codron, J.~Codron, A.~Baugh, and
  F.~Thackeray, 2005.
\newblock {Hominins, sedges, and termites: new carbon isotope data from the
  Sterkfontein valley and Kruger National Park}.
\newblock J. Hum. Evol. 48:301--312.

\bibitem[{Sponheimer et~al.(2006)Sponheimer, Loudon, Codron, Howells, Pruetz,
  Codron, de~Ruiter, and Lee-Thorp}]{Sponheimer:2006p718}
Sponheimer, M., J.~E. Loudon, D.~Codron, M.~E. Howells, J.~D. Pruetz,
  J.~Codron, D.~J. de~Ruiter, and J.~A. Lee-Thorp, 2006.
\newblock {Do ``savanna" chimpanzees consume ${\rm C_4}$ resources?}
\newblock J. Hum. Evol. 51:128--133.

\bibitem[{Stirling(1969)}]{Stirling:1969vf}
Stirling, I., 1969.
\newblock {Tooth wear as a mortality factor in the Weddell seal, {\it
  Leptonychotes weddelli}}.
\newblock J. Mammal. 50:559--565.

\bibitem[{Swennen et~al.(1983)Swennen, De~Bruijn, Duiven, Leopold, and
  Marteijn}]{Swennen:1983hfa}
Swennen, C., L.~De~Bruijn, P.~Duiven, M.~Leopold, and E.~Marteijn, 1983.
\newblock {Differences in bill form of the oystercatcher {\it Haematopus
  ostralegus}; a dynamic adaptation to specific foraging techniques}.
\newblock Neth. J. Sea Res. 17:57--83.

\bibitem[{Teaford and Oyen(1989)}]{Teaford:1989bc}
Teaford, M.~F. and O.~J. Oyen, 1989.
\newblock {Differences in the rate of molar wear between monkeys raised on
  different diets}.
\newblock J. Dent. Res. 68:1513--1518.

\bibitem[{Ungar et~al.(2008)Ungar, Grine, and Teaford}]{Ungar:2008tj}
Ungar, P.~S., F.~E. Grine, and M.~F. Teaford, 2008.
\newblock {Dental microwear and diet of the Plio-Pleistocene hominin {\it
  Paranthropus boisei}}.
\newblock PLoS ONE 3:e2044.

\bibitem[{Ungar et~al.(2012)Ungar, Krueger, Blumenschine, Njau, and
  Scott}]{Ungar:2012bp}
Ungar, P.~S., K.~L. Krueger, R.~J. Blumenschine, J.~Njau, and R.~S. Scott,
  2012.
\newblock {Dental microwear texture analysis of hominins recovered by the
  Olduvai Landscape Paleoanthropology Project, 1995-2007.}
\newblock J. Hum. Evol. 63:429--437.

\bibitem[{Ungar et~al.(2010)Ungar, Scott, Grine, and Teaford}]{Ungar:2010hk}
Ungar, P.~S., R.~S. Scott, F.~E. Grine, and M.~F. Teaford, 2010.
\newblock {Molar microwear textures and the diets of {\it Australopithecus
  anamensis} and {\it Australopithecus afarensis}}.
\newblock Philos. T. Roy. Soc. B 365:3345--3354.

\bibitem[{Ungar and Sponheimer(2011)}]{Ungar:2011em}
Ungar, P.~S. and M.~Sponheimer, 2011.
\newblock {The diets of early hominins}.
\newblock Science 334:190--193.

\bibitem[{Vincent(1985)}]{Vincent:1985hn}
Vincent, A.~S., 1985.
\newblock {Plant foods in savanna environments: a preliminary report of tubers
  eaten by the Hadza of Northern Tanzania.}
\newblock World Archaeol. 17:131--148.

\bibitem[{Vogel et~al.(2008)Vogel, van Woerden, Lucas, Utami~Atmoko, van
  Schaik, and Dominy}]{Vogel:2008ha}
Vogel, E.~R., J.~T. van Woerden, P.~W. Lucas, S.~S. Utami~Atmoko, C.~P. van
  Schaik, and N.~J. Dominy, 2008.
\newblock {Functional ecology and evolution of hominoid molar enamel thickness:
  {\it Pan troglodytes schweinfurthii} and {\it Pongo pygmaeus wurmbii}.}
\newblock J. Hum. Evol. 55:60--74.

\bibitem[{Weishampel et~al.(2004)Weishampel, Dodson, and
  Osm{\'o}lska}]{Weishampel:2004tpa}
Weishampel, D.~B., P.~Dodson, and H.~Osm{\'o}lska, 2004.
\newblock {The Dinosauria}.
\newblock University of California Press, Berkeley.

\bibitem[{White et~al.(2009)White, Asfaw, Beyene, Haile-Selassie, Lovejoy,
  Suwa, and Woldegabriel}]{White:2009p2395}
White, T.~D., B.~Asfaw, Y.~Beyene, Y.~Haile-Selassie, C.~O. Lovejoy, G.~Suwa,
  and G.~Woldegabriel, 2009.
\newblock {{\it Ardipithecus ramidus} and the paleobiology of early hominids}.
\newblock Science 326:64--64, 75--86.

\bibitem[{Williams et~al.(2005)Williams, Wright, Truong, Daubert, and
  Vinyard}]{Williams:2005dsa}
Williams, S.~H., B.~W. Wright, V.~d. Truong, C.~R. Daubert, and C.~J. Vinyard,
  2005.
\newblock {Mechanical properties of foods used in experimental studies of
  primate masticatory function}.
\newblock Am. J. Primatol. 67:329--346.

\bibitem[{Research(2012)}]{WolframResearchInc:2010ub}
Wolfram Research, 2012.
\newblock {Wolfram|KnowledgeBase}.
\newblock Wolfram Research, Inc, Champaign, Illinois.

\bibitem[{Wood and Constantino(2007)}]{Wood:2007cd}
Wood, B.~A. and P.~J. Constantino, 2007.
\newblock {{\it Paranthropus boisei}: Fifty years of evidence and analysis}.
\newblock Am. J. Phys. Anthropol. 134:106--132.

\bibitem[{Wood and Schroer(2012)}]{Wood:2012ct}
Wood, B.~A. and K.~Schroer, 2012.
\newblock {Reconstructing the diet of an extinct hominin taxon: the role of
  extant primate models}.
\newblock Int. J. Primatol. 33:716--742.

\bibitem[{Yamashita et~al.(2009)Yamashita, Vinyard, and Tan}]{Yamashita:2009fa}
Yamashita, N., C.~J. Vinyard, and C.~L. Tan, 2009.
\newblock {Food mechanical properties in three sympatric species of {\it Hapalemur} in
  Ranomafana National Park, Madagascar}.
\newblock Am. J. Phys. Anthropol. 139:368--381.

\bibitem[{Yeakel et~al.(2007)Yeakel, Bennett, Koch, and
  Dominy}]{Yeakel:2007p1410}
Yeakel, J.~D., N.~C. Bennett, P.~L. Koch, and N.~J. Dominy, 2007.
\newblock {The isotopic ecology of African mole rats informs hypotheses on the
  evolution of human diet}.
\newblock Proc. Roy. Soc. B 274:1723--1730.

\bibitem[{Youngblood(2004)}]{Youngblood:2004it}
Youngblood, D., 2004.
\newblock {Identifications and quantification of edible plant foods in the
  Upper (Nama) Karoo, South Africa}.
\newblock Econ. Bot. 58:S43--S65.

\end{thebibliography}

\newpage

\section*{Figures}

\newpage

\newpage

\newpage

\newpage

\newpage

\newpage

\begin{table}[tp]\footnotesize
\centering
\begin{tabular}{l>{\small} l >{\small} l>{\small} l>{\small} l}
\hline Parameter & Interpretation & Units & Value(s): Rich quality&Poor quality\\
\hline $X(t) = x$ & Energy reserves at time $t$ &10 [MJ]&  State Variable &\\ 
$V(t) = v$ & Enamel volume at time $t$ &100 [mm$^3$]& State Variable &\\
$K=k$ & Number of food items found &Count& Stochastic Variable &\\
$\Omega=\omega$ & Basal enamel wear &[mm]& Stochastic Variable &\\
\hline
$\gamma$ & Energetic Gain &10 [MJ]& $(1.5, 0.3, 1.6, 3.2)$ & $(1.4, 0.3, 1.4, 2.9)$ \\
$c$ & Energetic Cost (minimal) &10 [MJ]& $(0.7, 0.5, 0.7, 2.2)$ & $(1.1, 0.5, 0.7, 2.2)$ \\
  & Energetic Cost (maximal) &10 [MJ]& $(1.4, 1.2, 1.4, 2.8)$ & $(1.8, 1.2, 1.4, 2.8)$ \\
$\xi$ &Mean encounter rate &time$^{-1}$& $(3, 4, 3, 1)$ & $(2, 4, 3, 1)$\\
$\nu$&Dispersion & NA & $(3, 5, 3, 2)$  & $(2, 4, 3, 1)$ \\
$\eta$ & Digestibility &NA& $(0.9,0.7,0.8,0.9)$&Same \\
$A$ & Molar surface area &[mm$^2$]& $\sum^3_{m=1}\pi L^2_m$&Same \\
$b$ & Slope of enamel wear &[mm/k]& $0.0425$&Same \\
$E$ & Young's modulus &[MPa]& $(1, 10, 5, 200)$&Same \\
$R$ & Fracture toughness &[Jm$^{-2}$]& $(565, 330, 265, 1345)$&Same \\
$\bar{\omega}$ & Expected basal enamel wear & $\mu{\rm m}$ & 0.24 &Same \\
$\sigma$ & Basal enamel wear SD &$\mu{\rm m}$& 1.6 &Same \\
$d$ & Prob. of death at time $t$ & NA & ${\rm e}^{-10}$&Same \\
$Q(t)$ & Habitat quality at time $t$ & binary & {\rm r}&{\rm p} \\
\hline
$\bm \rho$ & Quality transition probability && Wet\hspace{10pt} $(0.8, 0.2; 0.2, 0.8)$ \\
  &matrix: $(\rho_{\rm rr}, \rho_{\rm rp}; \rho_{\rm pr}, \rho_{\rm pp})$& &Dry \hspace{7pt}  $(0.2, 0.8; 0.8, 0.2)$ &\\
  && &Auto. \hspace{0pt} $(0.8,0.2;0.8,0.2)$ &\\
\hline
$\Phi$ & Terminal fitness function ($t=T$)&  & \\
$F$ & Fitness function ($t < T$) &  & \\
$D^*(x,v)$ & Stationary decision matrix & & \\
$\hat{F}$ & Expected future fitness &  & \\
\end{tabular}
\caption{
Parameters and variables in the dynamic state variable model.
Parenthetical values (except for $\bm \rho$) are with respect to the foods: (fruit, grass leaves, USOs, arthropods).
Values for $E$ and $R$ are those when no mechanical advantage is included.
See methods for relevant references. Auto. = Autocorrelated.
}
\label{Tab:param}
\end{table}

\end{document}